\documentclass[12pt]{iopart}

\usepackage{graphicx}
\usepackage{caption}
\usepackage{subcaption}

\usepackage{csquotes}
\usepackage{url}
\usepackage{hyperref}
\usepackage{siunitx}
\usepackage{float}
\usepackage{multirow}

\usepackage{iopams}

\begin{document}

%Advancing Particle Simulation with Machine Learning through Normalizing Flows and Flow Matching Techniques

\title[End-to-end simulation of particle physics events $\dots$]{End-to-end simulation of particle physics events with Flow Matching and generator Oversampling}

\author{F Vaselli $\footnote[1]{Corresponding Author}$ $^{1, 2}$ , F Cattafesta $^{1, 2}$, P Asenov $^{2, 3}$ and A Rizzi $^{2, 3}$}

\address{$^1$ Scuola Normale Superiore, Pisa\\
$^2$ Istituto Nazionale di Fisica Nucleare, Pisa\\
$^3$ Universit\`a  di Pisa}
\eads{\mailto{francesco.vaselli@cern.ch}, \mailto{filippo.cattafesta@cern.ch}, \mailto{patrick.asenov.asenov@cern.ch}, \mailto{andrea.rizzi@cern.ch}}
\vspace{1pt}

\begin{indented}
\item[]February 2024
\end{indented}


\begin{abstract}
The simulation of high-energy physics collision events is a key element for data analysis at present and future particle accelerators. The comparison of simulation predictions to data allows looking for rare deviations that can be due to new phenomena not previously observed. We show that novel machine learning algorithms, specifically Normalizing Flows and Flow Matching, can be used to replicate accurate simulations from traditional approaches with several orders of magnitude of speed-up. The classical simulation chain starts from a physics process of interest, computes energy deposits of particles and electronics response, and finally employs the same reconstruction algorithms used for data. Eventually, the data are reduced to some high-level analysis format. Instead, we propose an end-to-end approach, simulating the final data format directly from physical generator inputs, skipping any intermediate steps. We use particle jets simulation as a benchmark for comparing both \emph{discrete} and \emph{continuous} Normalizing Flows models. The models are validated across a variety of metrics to identify the most accurate. We discuss the scaling of performance with the increase in training data, as well as the generalization power of these models on physical processes different from the training one. We investigate sampling multiple times from the same physical generator inputs, a procedure we name \emph{oversampling}, and we show that it can effectively reduce the statistical uncertainties of a dataset. This class of ML algorithms is found to be capable of learning the expected detector response independently of the physical input process. Their speed and accuracy, coupled with the stability of the training procedure, make them a compelling tool for the needs of current and future experiments.
\end{abstract}


\section{Introduction}\label{sec1}
The simulation of high-energy physics (HEP) events is a necessary and complex task involving various steps. Comparing simulated predictions with actual data helps to notice unusual variations that might signal new, previously unseen phenomena or allows measuring processes known for being extremely rare. At the Large Hadron Collider (LHC), billions of simulated collision events are necessary, and this number is expected to increase with the development of future detectors and accelerators \cite{Software:2815292}. The production of simulated samples usually starts from the outputs of a physics process generator (in the following referred to as \emph{gen}) such as \textsc{pythia} \cite{bierlich2022comprehensive}, describing the list of particles resulting from the physical process at the collision point, and it ends with some analysis-ready data.

After gen, the next step, called simulation (or \emph{sim}), based on frameworks, such as \textsc{geant4} \cite{AGOSTINELLI2003250}, relies on numerical Monte-Carlo approaches to propagate the particles through the detector and compute the energy deposits into the particle sensors. This step is followed by the conversion of energy deposits into digital electronics readout signals (\emph{digi}) reproducing the actual lowest level output of the detector during data taking. Finally, the electronics signals are converted back to physics quantities with the very same reconstruction (\emph{reco}) algorithms used for real detector data.

\begin{figure}[ht]
    \centering
    \includegraphics[width=0.7\columnwidth]{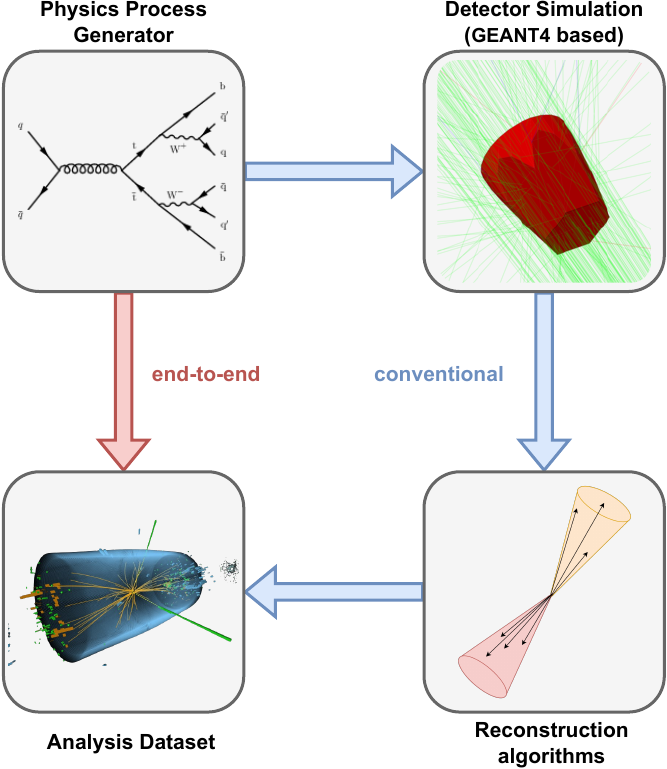}
    \caption{The typical simulation chain for HEP collaborations, consisting of \emph{generation}, \emph{simulation} \& \emph{digitization}, \emph{reconstruction} and \emph{analysis data reduction} steps. The present work investigates the capabilities of Normalizing Flows models to do end-to-end simulation from the first step directly to the last one of this chain (figures taken from \cite{cms_eventdisplay,enwiki:1195481216}).}
    \label{fig:problem_state}
\end{figure}

This conventional simulation chain solves the problem of sampling from $P(\rm reco|\rm {gen})$, where the analytical form of this probability density function (\emph{pdf}) is unknown, and the various steps parameterize different contributions:

\begin{equation}
    P(\rm reco|\rm {gen}) = P_{\rm {sim}}(sim|\rm {gen}) \times P_{\rm {dig}}(digi|\rm {sim}) \times P_{\rm {rec}}(reco|\rm {digi}).
\end{equation}

Given that the reconstruction step is common in actual data and simulation, its output is available in both and allows the data to simulation comparisons that are the key ingredients for HEP analysis. The data analysis typically starts with various data reduction steps where the physics information is further transformed into higher level observables, for example combining information from multiple detectors or multiple particles into some derived quantities. The output of these additional steps is usually in a much simpler format than the one at the \emph{reconstruction} level.

This type of approach serves as the backbone of the majority of simulation frameworks for the various HEP collaborations, and it has been proved to be accurate and adaptable to the various experimental scenarios. However, the time and computational resources taken by this procedure are substantial, and are expected to increase rapidly as we reach new frontiers of energy and luminosity.

Several approaches exist to mitigate this issue through the development of faster simulation frameworks. Examples are \emph{Delphes} \cite{de_Favereau_2014}, a framework for fast simulation of a generic collider experiment, and various experiment-specific toolkits. Delphes uses parametric smearing on the generator level information, going \emph{end-to-end} by directly producing analysis-ready data.  The application of ML to such an approach for analysis-specific applications is outlined in \cite{chen2020data}.

In the present work, we investigate the use of Normalizing Flows (NF), a type of generative machine learning (ML) algorithms, for performing \emph{end-to-end} simulation, similar to Delphes, but with a higher degree of accuracy. Specifically, we focus on optimizing the choice of the NF model based on physics-aware metrics.
As in the traditional simulation chain, we use random noise \emph{z} as input, along with generator outputs (gen). The use of gen in input is called \emph{conditioning}, as this information conditions the output of our model according to the different physical processes $P(\rm reco|\rm {gen})$.
The idea is to go from the gen inputs to the final event description directly by learning a surrogate $f_{\phi}$ of the \emph{pdf} from datasets obtained with conventional simulations: $f_{\phi}(\rm event|\rm{gen}) \approx P_{\rm{Target}}(\rm event|\rm{gen})$, where $\phi$ are the parameters of a Neural Network. In this work, we demonstrate that such an approach can retain a good amount of details, and it faster than conventional simulation, especially when accelerated with GPUs.
The general problem of HEP simulation and the proposed approach are illustrated in Figure \ref{fig:problem_state}.

The usage of ML techniques for simulation has already been proposed in HEP. For a comprehensive review, we point the reader to \cite{Butter_2023}. As examples, the CMS FastSim simulation toolkit \cite{Giammanco_2014} is already using ML for both parametrization of single steps and refinement of final results \cite{osti_2202826}. The LHCb ultra-fast simulation framework Lamarr \cite{barbetti2023lamarr} revolves around the use of Generative Adversarial Networks. Generally speaking, ML models are often applied to single steps of the simulation chain, rather than in an end-to-end fashion: an example is the simulation of calorimeter response as done in \cite{buhmann2024caloclouds, ernst2023normalizing}. Often times, existing approaches make use of more popular generative algorithms, such as Generative Adversarial Networks or Variational Autoencoders, rather than Normalizing Flows. Some convincing applications of NF have nonetheless been presented, for example by the ATLAS Collaboration in the context of photon simulation \cite{xu2023generative}, from researchers studying anomaly detection at the LHC \cite{Jawahar_2022} and for usage within physical generators \cite{heimel2023madnis}. A general study on the robustness of NF has been presented in \cite{coccaro2024comparative}.  However, usage of Flows is often limited to analysis/process-specific applications. 
The use of newer architectures includes the exploration of Diffusion Models for jet simulation, as discussed in \cite{Mikuni_2023}.

In this work, we present and discuss multiple NF models on a toy-dataset of particle jets, with a focus on assessing the most accurate class of models. The main contributions of this work are the following.

\begin{itemize}
    \item We investigate both \emph{discrete} NF, where the final transformation $f_{\phi}(\rm event|\rm {gen})$ is made up of multiple, simple discrete transforms; and \emph{continuous} NF, where the model learns a \emph{vector field}, which is used in computing the trajectories of an ordinary differential equation (ODE) which maps noise to data. We introduce a set of physically-motivated metrics, and we use them to compare the models.
    \item We compare speed and accuracy for the better performing models. We investigate if models trained on a small dataset can be used to produce a bigger dataset without introducing strong biases. We also show the generalization power of these algorithms, testing them on physical processes not seen during training.
    \item Finally, considering that the speed of these types of algorithms is even greater than that of physical generators, we discuss the possibility of producing multiple simulations starting from the same generator information \emph{gen}, a procedure we name \emph{oversampling}. As far as we know, this is the first time that such an approach has been proposed and discussed in the context of HEP simulation. We introduce a statistical procedure for handling histograms with oversampled data.
\end{itemize}

\section{Related work}\label{sec2}
This work focuses on the optimization of Flow algorithms and has been performed on a toy-dataset mimicking an HEP experiment. This effort is related to the work within the CMS Collaboration to use these techniques on actual experiment datasets \cite{Vaselli:2858890}. We expect that the outcome of our optimization will be a useful input to improve the CMS application of Normalizing Flow for end-to-end simulation.

Several works about ML approaches to simulation in HEP have already been presented over the years. Those employing Normalizing Flow explore topics such as fast calorimeter simulation \cite{Krause_2023, krause2023caloflow}, particle cloud simulation \cite{buhmann2023epicly}, jet  constituents simulation \cite{Bellagente_2020, birk2023flow}, and high-level event simulation \cite{butter2023jet, Butter_2023_1, Gao_2020, gavranovič2023systematic, käch2022jetflow}.

\section{Methods}\label{sec3}

Normalizing Flows (NF) are a class of generative machine learning models, which aim to express the unknown data \emph{pdf} $P(x)$ starting from a simple base distribution $B(z)$ (usually a Gaussian distribution) through an invertible mapping $f_{g}$, $x = f_{g}(z),\; z = f_{g}^{-1}(x)$. We refer to \cite{papamakarios2021normalizing} for a comprehensive review of existing NF algorithms.

\subsection{Continuous Normalizing Flows}
\emph{Continuous} flows are a type of flows which maps the base distribution into the target one through a \enquote{continuous} transformation parametrized by a variable \emph{t} $\in [0, 1]$, such that $P_{t=0}(z) = B(z)$ (the base noise distribution) and $P_{t=1}(z) = P(x|g)$ (the unknown data distribution). At each given \emph{t}, the current flow transformation is fully specified by a \emph{vector field} $v_{t, g}$:

\begin{equation}
\frac{d}{dt} f_{t,g}(z) = v_{t,g}(f_{t,g}(z)), \quad f_{0,g}(z) = z \; \rm {(Identity)}
\end{equation}

and the samples $x = f_{t=1,g}(z) = \int_0^1 dt\, v_{t,g}$ are obtained by integrating this ODE.

It has been observed \cite{dax2023flow} that continuous NF have the advantage that the network is less constrained by the specific choice of a transformation and can be more expressive than discrete ones. Additionally, this type of transformation acts simultaneously on all the features, while for discrete flows we need to introduce coupling/autoregressive mechanisms. However, training this class of flows traditionally relied on many passes through the network to solve the ODEs, and it is often simpler to train discrete flows.

\subsection{Flow matching}
A possible alternative to train continuous NF is the use of \emph{Flow matching} \cite{lipman2023flow}. This novel approach consists in casting the training into a regression problem, which is simpler to address. The aim is to learn the vector field $v_{t, g}$ as the output of the network. Because it is unknown, another vector field $u_{t}$ is taken as the regression target, which defines a \emph{probability path} $P_{t}$ running from $B(z)$ to $P(x|g)$ as \emph{t} increases. If $P_{t}$ does this mapping well, then $u_{t}$ is close to what we would like to learn, i.e. $v_{t, g}$. The intuition behind \cite{lipman2023flow} is that $u_{t}$ and $P_{t}$ can be constructed in a \emph{sample-conditional} basis, i.e., depending on the training sample $x$. The loss for the parameters $\phi$ of the network then becomes:

\begin{equation}
    \mathcal{L}_{\rm {CMB}}(\phi) = \mathbb{E}_{t\sim\mathrm{Unif}[0,1], x\sim P(x|g), z_t \sim B_t(z|x)} \left[ \|v_{t,g}(z_t| \phi) - u_t(z_t|x)\|^2 \right].
\end{equation}

which is a simple regression loss. We can construct various types of probability paths and their associated vector fields $u_{t}$. In this work, we draw from recent developments in Flow matching presented in \cite{tong2023improving}, and distinguish between two main classes of paths:

\paragraph{Target conditional flow matching} This is the flow matching originally proposed by \cite{lipman2023flow}. Let \emph{x} be a single training data sample. It defines the probability path and the associated vector field as:

\begin{equation}
    P_t(z|x) = \mathcal{N}(z | tx, (t\sigma_{\rm {min}} - t + 1)^2),
    \label{flow-matching}
\end{equation}
\begin{equation}
    u_t(z|x) = \frac{x - (1 - \sigma_{\rm {min}})z}{1 - (1 - \sigma_{\rm {min}})t},
\end{equation}

which is a probability path from the standard normal distribution $(B(z) = P_0(z|x) = \mathcal{N}(z; 0, I))$ to a Gaussian distribution centred at \emph{x} with standard deviation $ \sigma_{\rm {min}}$, $(P_1(z|x) = \mathcal{N}(z; x,  \sigma_{\rm {min}}^2))$.

\paragraph{Basic form of conditional flow matching} This is the I-CFM discussed in \cite{tong2023improving}. In the I-CFM, we identify \( x \) with a pair of random variables, a source sample \( x_0 \) (the noise) and a target sample \( x_1 \) (the training data). We let the paths be Gaussian flows between \( x_0 \) and \( x_1 \) with standard deviation \( \sigma_{\rm {min}} \), defined by:
\begin{equation}
P_t(z|x) = \mathcal{N}(z | t x_1 + (1 - t) x_0, \sigma_{\rm {min}}^2),
\end{equation}
\begin{equation}
u_t(z|x) = x_1 - x_0.
\end{equation}

We note that in this formulation there is no assumption on the base distribution $B(z)$ to be Gaussian, as was for the previous class of paths. This means that we can use I-CFM to create NF starting from other base distributions, such as the Uniform one.

\subsection{Particle Jets dataset}
We study the simulation of particle jets originating in proton-proton collisions. Note however that the approach presented in this work could be adapted to the simulation of any arbitrary physical objects and distributions. By combining the simulation of multiple objects, we can simulate end-to-end an entire event.

The ultimate aim of the approach presented is to learn the response function of detectors as accurately as that simulated with \textsc{geant4} based toolkits. However, in this paper, we employed a toy-dataset for the comparison of various flow models. The \textsc{geant4} simulation, which would be the realistic target, is replaced with Delphes-like smearing of generator level quantities. In order to make the dataset more challenging for the NF models, we introduced several correlations and dependencies that are expected in a real detector, e.g. the dependency of momentum measurement bias and resolution as a function of the jet energy or the correlation of such a resolution with the jet flavour. More details on the dataset and on the various processes simulated are given in \ref{secA1}.

Eventually, the dataset we built consists of pairs of generator-level (gen) jets and associated reconstructed (reco) jets 
 similar to those obtained after the full simulation chain of particle-matter interaction, digitization and reconstruction algorithms.

We created multiple datasets starting from the \textsc{pythia} generator, in proton-proton collisions for 4 different physical processes: top quark pair production ($pp \rightarrow t\bar{t}$), Z boson production in association with jets ($pp \rightarrow Z$+Jets), W boson pair production ($pp \rightarrow WW$) and multiple jets production ($pp \rightarrow JJ + X$). We clustered stable final state particles in order to define \emph{generator-level} jets through the \emph{FastJet} \cite{FastJetmanualCacciari_2012} library (specifically the \emph{anti-}$k_T$ jet clustering algorithm \cite{antiktCacciari_2008}, with $R = 0.4$). We kept only jets with $p_T>15$ GeV (we use the HEP experiment convention, with Lorentz 4-vectors in cylindrical coordinates and the polar angle $\theta$ replaced by the \emph{pseudorapidity} $\eta = -\ln[\tan(\theta/2)]$).

This strategy has the advantage of producing a target dataset complex enough to showcase the capabilities of flow-based simulation at a fraction of the complexity and computing cost than that of running a full simulation process.

In the following section, we list the most notable features of our dataset, emphasizing non-trivial distributions and correlations introduced by our toy-physics simulation, which we would like our models to reproduce correctly. Table \ref{table:datasets} shows the list of the 6 gen-level features which we give as input to our models (the conditioning on the \emph{pdf}) and the two different sets of reco target variables: a basic one with 5 variables, and the extended one with 16 variables.

\begin{table}
\caption{The two datasets used in this work: one with 6 input generator-level variables and 5 target reco-level variables; an extended one with the same inputs and 16 target reco variables in total.}

\resizebox{\textwidth}{!}{
\begin{tabular}{@{}lll}
\br
{\bf Generator level variables } & {\bf  Description }\\
\mr
$p_T$, $\eta$, $\phi$, mass & Kinematic properties of the generated jet \\
jet flavour & Distinguishing b, c jets from light quarks or gluon jets \\
number of $\mu$ in jet & Counting the number of muons within the jet radius \\
\mr
{\bf Basic reconstructed variables }&{\bf Description }\\
\mr
$p_T$, $\eta$, $\phi$, mass & Kinematic properties of the reconstructed jet \\
b-tagging discriminator & Score in [0,1] mimicking a tagging algorithm  \\
number of constituents & Counting the number of reconstructed jet constituents \\
\mr
{\bf Extended dataset variables } & {\bf Description (in addition to basic variables) }\\
\mr
Neutral Hadron Fraction (nhf) & fraction of jet energy carried by Neutral Hadrons \\
Charged Hadron Fraction (chf) & fraction of jet energy carried by Charged Hadrons \\
Neutral Electromagnetic Fraction (nef) & fraction of jet energy carried by photons and $\pi^0$ mesons \\
Charged Electromagnetic Fraction (cef) & fraction of jet energy carried by electrons \\
Quark-Gluon discriminator (qgd) & Discriminator score mimicking a quark/gluon tagging algorithm  \\
Jet Identification (jetId) & Discriminator score mimicking a jet Identification algorithm \\
Number of Charged Particles (ncharged) & Number of reconstructed charged particles\\
Number of Neutral Particles (nneutral) & Number of reconstructed neutral particles \\
c-tagging discriminator & Score of c-tagging algorithm, correlated with b-tagging  \\
Number of Secondary Vertices (nSV) & Poisson distributed number of Secondary Vertices in jets\\
\br
\end{tabular}}
\label{table:datasets}
\end{table}

\begin{figure}
    \centering
    \includegraphics[width=0.5\linewidth]{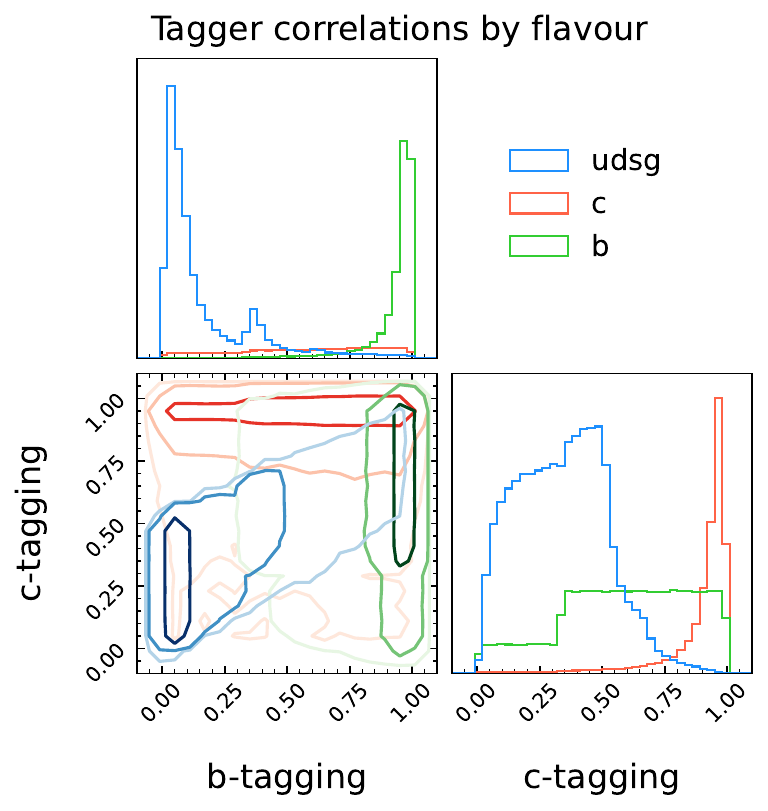}
    \caption{The correlations between the two target tagging distribution are shown as contours lines in the bottom left panel, coloured by the flavour of the jets. The two target distribution have non-trivial correlations for light jets,  and b or c flavours correspond to peaks in the tagging distributions.}
    \label{fig:tagging-corrs}
\end{figure}

Among them, we emphasize the presence of pseudo \emph{b-tagging} and \emph{c-tagging} scores, which have been defined to simulate the performance of generic tagging algorithms. Their value is bounded between 0 and 1, where the jet flavour of the \emph{b} quark is associated with higher b-tagging scores, but also to generally higher values in the c-tagging score, as the two quarks produce similar features in the jets which can fool these types of algorithms. Light quarks and gluons are instead associated with lower scores. We show in Figure \ref{fig:tagging-corrs} the 1-d distributions and correlations between the b-tagging score and c-tagging score based on the jet flavour.

Another correlation we introduced in the dataset is the one between the gen-level input $p_T$ and the reconstructed one. If we consider the $p_T$ response (ratio of the reco to the gen), its resolution shrinks as the gen $p_T$ increases.

Variables are preprocessed before training. Specifically, we \emph{standardized} both input and target features by subtracting from each one its mean and dividing it by its standard deviation. Integer target variables were smeared with random uniform noise in [-0.5, 0.5], a process known as \emph{dequantization}. Categorical variables, such as the jet flavour, were \emph{one-hot} encoded into a series of 0/1 flags.

\subsection{Validation metrics}
We selected 6 different kinds of metrics to evaluate our models. The metrics chosen are the following.
\begin{itemize}
    \item The 1-d \emph{Wasserstein} score (WS) \cite{Kansal_2023} and the two-sample \emph{Kolmogorov-Smirnov} distance (KS) for comparing 1-d distributions between the target and the samples produced by the model. A WS is assigned to each variable.
    \item The \emph{Fréchet} distance as a global measure. It is the distance between Multivariate Gaussian distributions fitted to the features of interest, which \cite{Kansal_2023} calls the Fréchet Gaussian Distance (FGD). It is generally called the Fréchet Inception Distance (FID) in image generation tasks:
    \begin{equation}
    d^2(x, y) = \|\mu_x - \mu_y\|^2 + \Tr(\Sigma_x + \Sigma_y - 2(\Sigma_x \Sigma_y)^{1/2}).
    \end{equation}
    \item \emph{Covariance matching}: another global metric used to measure how well an algorithm is modelling the correlations between the various target features. Given the covariance matrices of the two samples, target and model, we compute the \emph{Frobenius Norm} of the difference between the two: 
    \begin{equation}
    || \mathrm{Cov}(X_{\rm{target}}) - \mathrm{Cov}(X_{\rm{model}}) ||_F = \sqrt{\sum_{i=1}^{m}\sum_{j=1}^{n}|c^{\mathrm{t}}_{ij} - c^{\mathrm{m}}_{ij}|^2}. \\
    \end{equation}
    Correlations in the model samples are also visually evaluated through the use of dedicated plots.
    \item As b and c-tagging are such important tasks in the study of jets, we compute the \emph{receiver operating characteristic} (ROC) curves for both scores. To quantify the performance of a model, we compute the difference in log-scale between the ROC coming from the model and that from the target distribution. Log-scale is used because the true positive rate (TPR) and false positive rate (FPR) span different orders of magnitude. We call this evaluation metric the \emph{Area Between the Curves} (ABC).
    \item Finally, we implement a \emph{classifier two-sample test} (c2st): we train a classifier to distinguish between training samples and samples coming from our models, giving as additional input the gen information. The output is the percentage $P_{\rm{c2st}}$ of samples which were \emph{incorrectly} classified. For the optimal model, it has a maximum value of 0.5. We thus report our results as $0.5 - P_{\rm{c2st}}$: in this way the best model has the lowest c2st value. We use a scikit-learn \cite{scikit-learn} \emph{HistGradientBoostingClassifier} with default parameters as our classifier.
\end{itemize}

\subsection{Training of models}
We use the PyTorch \cite{paszke2019pytorch} package, and the derived torchcfm \cite{tong2023improving,tong2023simulation} package, for creating and training the models. Additionally, the dingo package \cite{dax2023flow} for analysing gravitational wave data has been a major source of inspiration when designing the models. Models are trained on \num{500}k training samples, then compared and validated on a separate \num{200}k test split. The training and validation data are generated from the $t\bar{t}$ process. This process has wide tails in its distributions, making it useful for training the models to capture the correct detector response.  We use Gaussian noise as our base distribution, and for flow matching training routines we typically set $\sigma_{\rm {min}} = 10^{-4}$ (see Equation \ref{flow-matching}). We train for \num{1000} epochs. More details about the various models architectures, hyperparameters and naming conventions are available in \ref{secA2}. The code for reproducing the datasets along with training and validation of the best model is released \href{https://github.com/francesco-vaselli/FlowSim}{here}\footnote[1]{https://github.com/francesco-vaselli/FlowSim}. 

\begin{figure}
    \centering
    \begin{subfigure}[t]{0.45\columnwidth}
        \centering
        \includegraphics[width=\linewidth]{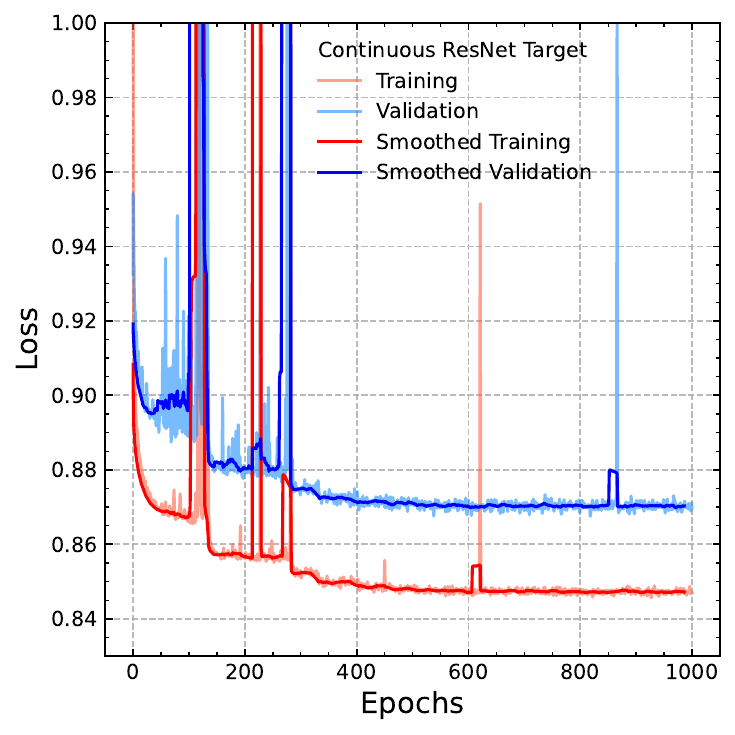}
        \caption{}
        \label{fig:losses_crt}
    \end{subfigure}
    \hfill 
    \begin{subfigure}[t]{0.45\columnwidth}
        \centering
        \includegraphics[width=\linewidth]{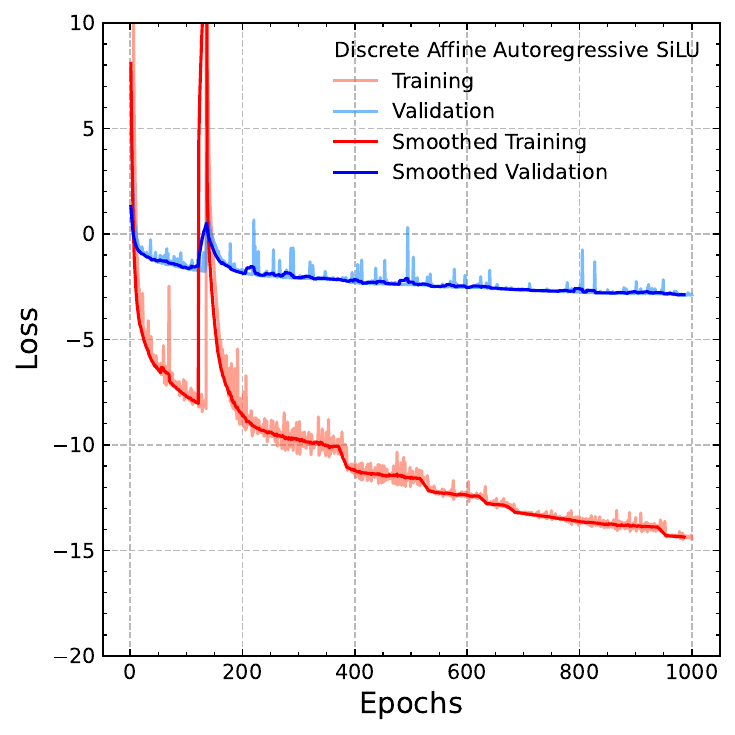}
        \caption{}
        \label{fig:losses_ac}
    \end{subfigure}
    \caption{Model losses for CRT and DAC. For Continuous models on the left (see CRT, Table \ref{tab:gencomp}), and for Discrete models on the right. We also report the smoothed loss values across a window of 15 epochs. (a) Continuous models: The loss does not show signs of overfitting. Sharp drops are due to the reduction of the learning rate on plateaus. (b) Discrete models: The loss shows a widening gap between Training and Validation, suggesting a possible overfitting.}
    \label{fig:losses}
\end{figure}

For the continuous models, we observe absence of overfitting. As an example, we show in Figure \ref{fig:losses} the loss for a continuous and a discrete model.

\subsection{Initial model comparison}
We explored different configurations of discrete and continuous NF models. In order to compare them, we computed 55 different metrics across the different target variables. We aggregate Wasserstein and KS metrics by performing a mean across variables.

The main models compared were:

\begin{itemize}
    \item Continuous ResNet Target (CRT) a continuous flow where a ResNet architecture is learning the vector field, trained in a Target Flow Matching regime;
    \item Continuous ResNet Basic (CRB), same as before but with a Basic Flow Matching;
    \item Continuous MLP Basic (CMB), same as before but with a MLP architecture;
    \item Discrete Affine Autoregressive (DAA), a discrete flow with affine transform and autoregressive conditioner, see \cite{papamakarios2021normalizing};
    \item Discrete Affine Coupling (DAC), a discrete flow with affine transform and coupling conditioner.
\end{itemize}

\begin{figure}
    \centering
    \begin{subfigure}[t]{0.45\columnwidth}
        \centering
        \includegraphics[width=\linewidth]{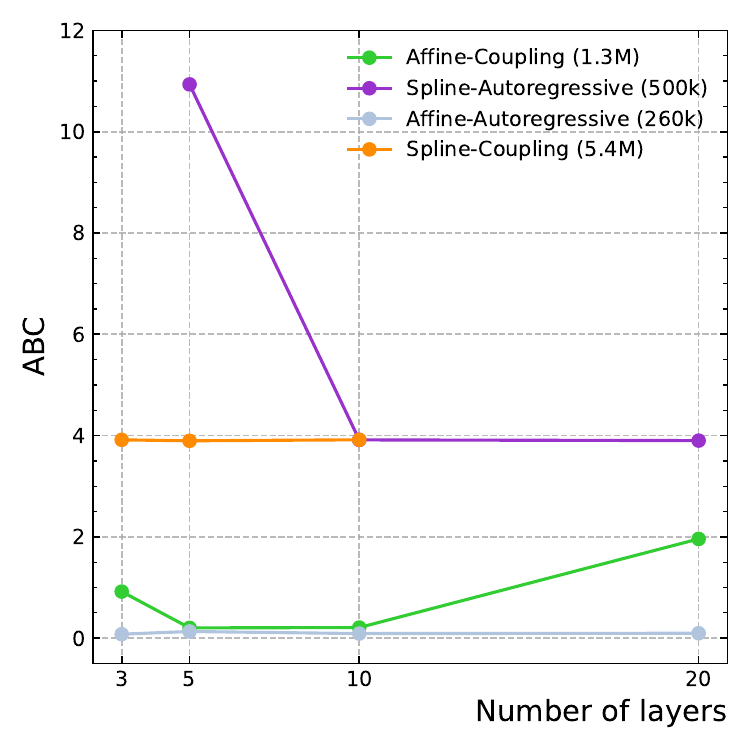}
        \caption{}
        \label{fig:discrete_abc}
    \end{subfigure}
    \hfill 
    \begin{subfigure}[t]{0.45\columnwidth}
        \centering
        \includegraphics[width=\linewidth]{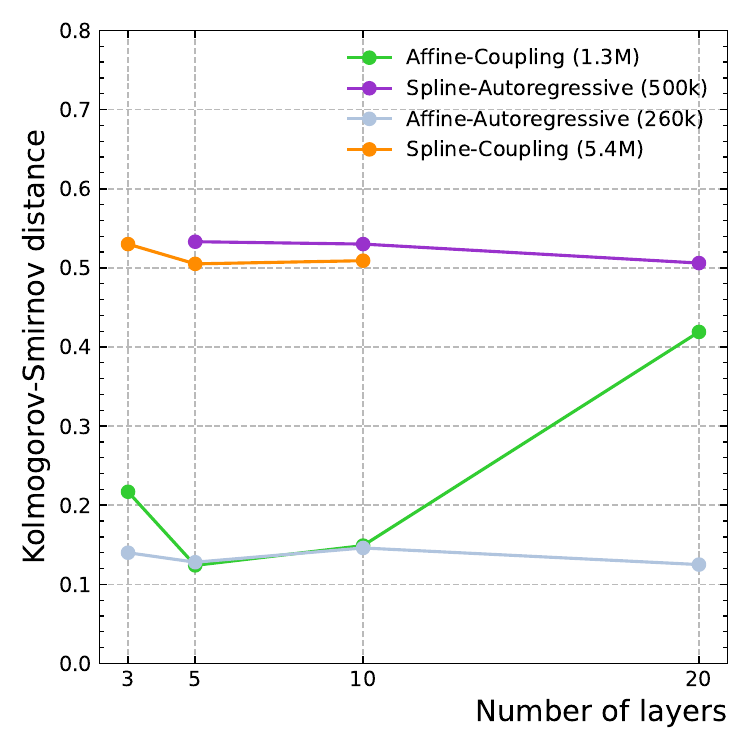}
        \caption{}
        \label{fig:discrete_ks_mean}
    \end{subfigure}
    \caption{Analysis of discrete models on the small dataset, showcasing the performance of Affine Coupling 5 layers and Affine Autoregressive 20 layers in both the ABC (left) and KS mean (right) metrics versus the other classes of models, computed as the median over the last 100 epochs. For each model, we report the number of trainable parameters in parentheses.}
    \label{fig:discrete_comp}
\end{figure}

We first performed a scan on the basic dataset, as we had many possible configurations of hyperparameters to test, especially in the discrete case. We used those results to perform a smaller set of informed trainings on the extended dataset. In particular, Figure \ref{fig:discrete_comp} shows the behaviour of the median over the last 100 epochs for global metrics such as the ABC and KS mean as a function of the different numbers of transformation layers in autoregressive and coupling flows. The Affine Coupling 5 layers and Affine Autoregressive 20 layers emerge as the best family of discrete architectures and have been retrained on the extended dataset along with the continuous flows models for our final comparison.

\begin{figure}
    \centering
    \includegraphics[width=\linewidth]{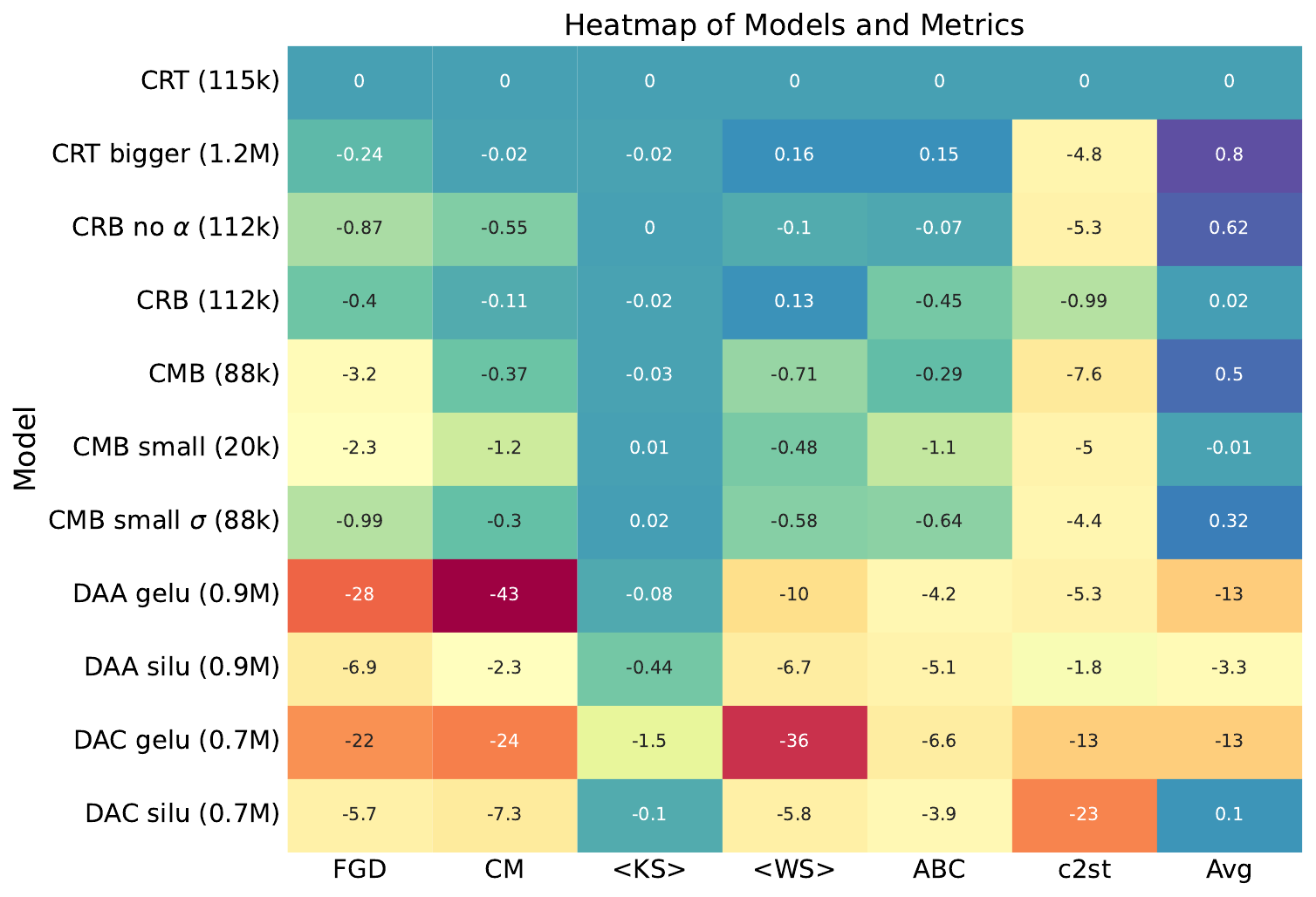}
    \caption{This plot quantifies how far each model performance is from the CRT baseline one. We compute the distance between the metrics of the model and the baseline one, and then divide by the sum in quadrature of the deviations. By averaging results across all metrics, we get the average \emph{significance}, telling us how many deviations there are between two models. Negative values indicate lower performance, positive ones stand for a better one. The number of trainable parameters for each model is reported in parentheses.}
    \label{fig:heatmap}
\end{figure}

For the extended case, we computed the metrics for each model every 10 epochs and their median on the last 10 evaluations. As the differences in performances for some models are close to the epoch-to-epoch oscillations, we computed the \emph{significance} $\Delta_{\rm{metric}}/\sqrt{\sigma_{\rm{diff}}^2 + \sigma_{0}^2}$ where $\Delta_{\rm{metric}}$ is the difference between the median values between the model and the reference, $\sigma_{\rm{diff}}^2$, is the variance of the difference and $\sigma_{0}$ is the measured value of the metric when comparing two simulations of the same gen data. The CRT model has been selected as reference, being the better performing one, as seen in Figure \ref{fig:heatmap} which shows the comparison across all the metrics. In particular, the last column shows the average of the metrics, and we can observe that the CRT, CRT bigger and CRB models are the best performing and very close to each other.

We also notice that continuous flows consistently achieve better results than discrete ones. Furthermore, they do so using a much smaller number of parameters, at least an order of magnitude lower. The \enquote{CMB small} model outperforms all discrete flows, despite having less than \num{20}k parameters compared to almost a million for the latter ones.

\begin{figure}[ht]
    \centering
    \includegraphics[width=0.5\columnwidth]{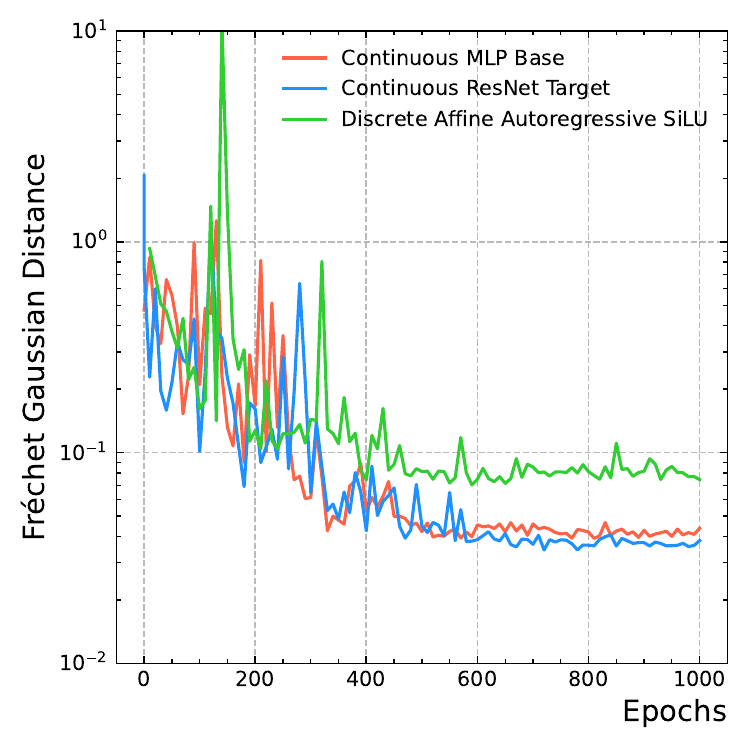}
    \caption{A comparison between models: the behaviour of the FGD as a function of training epochs. Continuous models converge to lower values.}
    \label{fig:fgd global}
\end{figure}

\begin{figure}[ht]
    \centering
    \begin{subfigure}[t]{0.45\columnwidth}
        \centering
        \includegraphics[width=\linewidth]{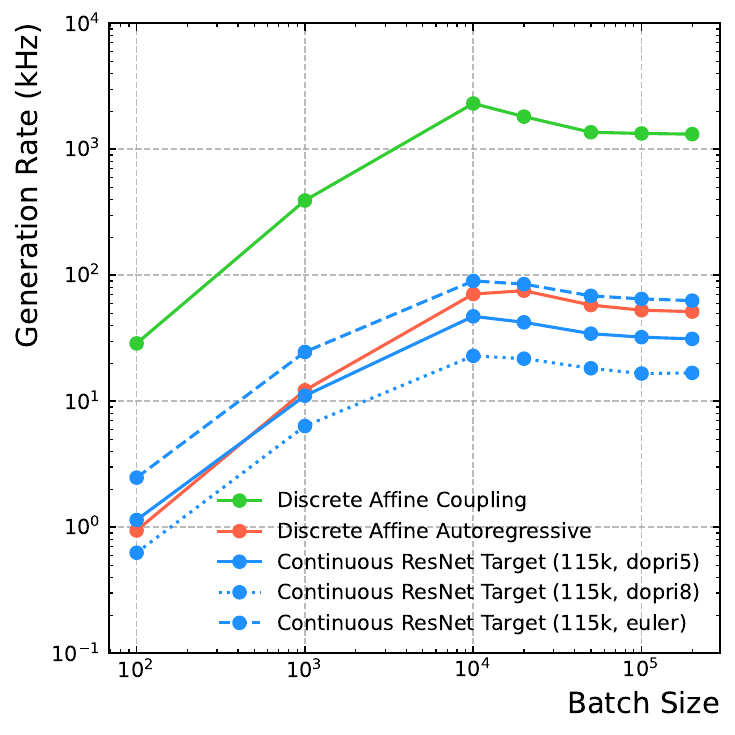}
        \caption{}
        \label{fig:timing}
    \end{subfigure}
    \begin{subfigure}[t]{0.45\columnwidth}
        \centering
        \includegraphics[width=\linewidth]{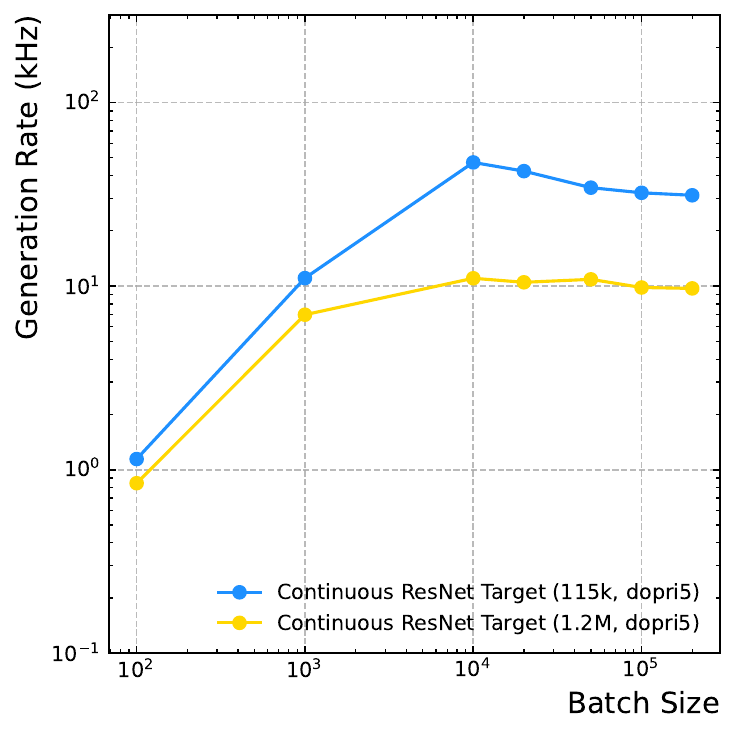}
        \caption{}
        
    \end{subfigure}
    \caption{The rate of sampling one batch of a given size for different models. On the right (b), a specific look at continuous model architectures of varying sizes demonstrates the impact of increased parameters on the generation rate.}
    \label{fig:timing_global}
\end{figure}

A comparison between different models during training is illustrated in Figure \ref{fig:fgd global}, which shows the behaviour of the FGD metric for different models as a function of the epoch. We can see how continuous models converge faster and to smaller values compared to the discrete one.

Another important factor to take into account is the generation (or \emph{sampling}) speed of the different types of models. We compare in Figure \ref{fig:timing_global} the CRT, DAA and DAC models speed by sampling with different batch sizes on a single GeForce RTX 4060 NVIDIA GPU. As expected, the DAC models are the fastest by a wide margin, thanks to the use of coupling layers. On the other hand, the CRT performance is in a similar range of the one of DAA models. For continuous models, sampling speed is strongly influenced by the choice of the ODE solver. To show this, we tested our models with the Dormand-Prince of Order 5 (\emph{dopri5}), Dormand-Prince of Order 8 (\emph{dopri8}) and Euler of Order 1 (\emph{euler}) solvers, with different absolute and relative tolerances, and we found them to have compatible results. For the simplest solver, the CRT model is as fast as the Discrete Autoregressive one. The size of the models also plays a role, as seen in Figure \ref{fig:timing_global}, showing the CRT and CRT bigger sampling speed side by side to showcase the impact of model size on the generation speed. To put the speed of our approach into context, we should consider that a single particle physics event has tens of objects to be simulated, and conventional simulation approaches can take tens of seconds to produce a single event. It is then clear that a multiple kHz speed on single objects translates into orders of magnitude speed-up with respect to conventional approaches, see Section \ref{sec:app}.

\section{Results}\label{sec4}
In the following, we show results obtained by sampling from a separate test split of \num{650}k jets for the best model (CRT). We used the dopri5 solver, with absolute and relative tolerances of $10^{-5}$.

\begin{figure}[H]
    \centering
    \begin{subfigure}[t]{0.45\columnwidth}
        \centering
        \includegraphics[width=\linewidth]{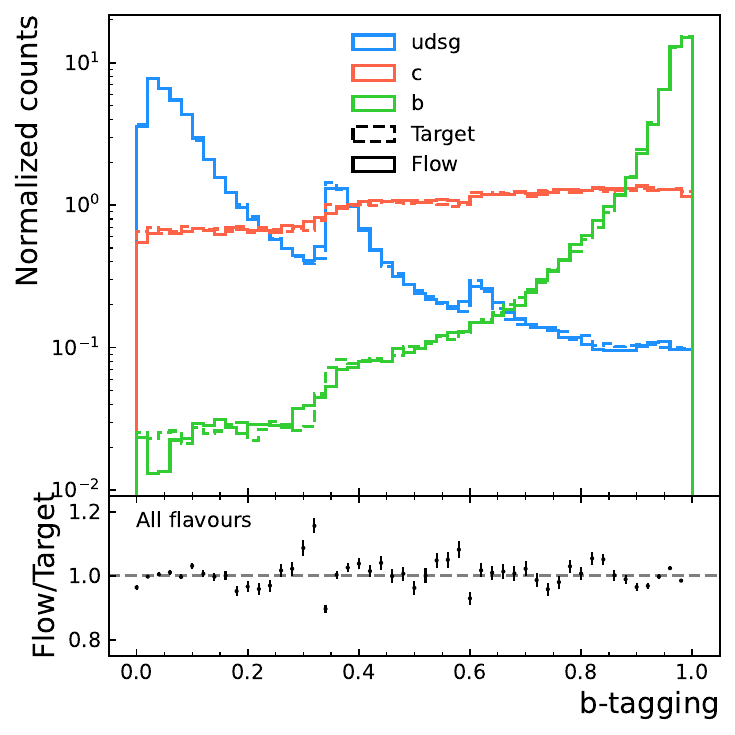}
        \caption{}
        \label{fig:ttbar_btag}
    \end{subfigure}
    \begin{subfigure}[t]{0.45\columnwidth}
        \centering
        \includegraphics[width=\linewidth]{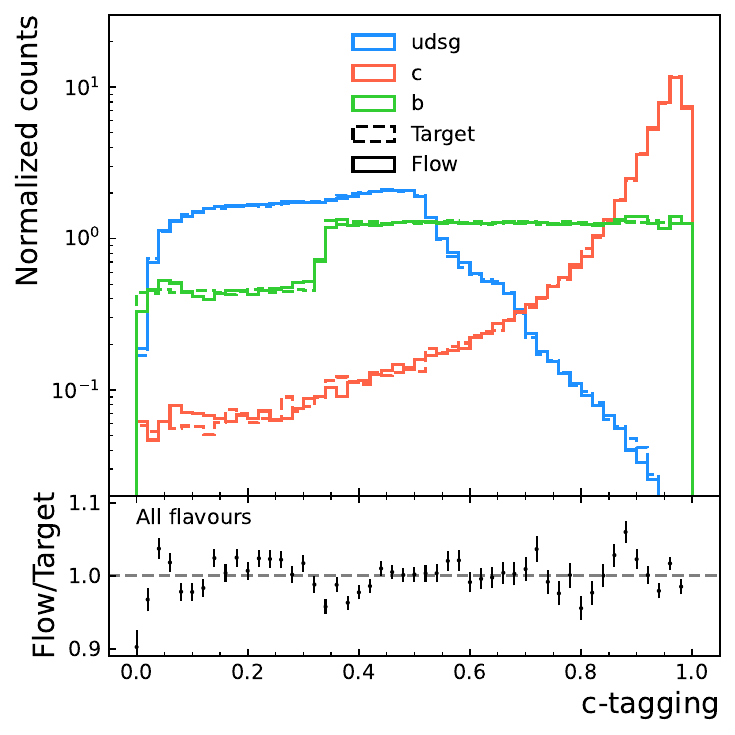}
        \caption{}
        \label{fig:ttbar_ctag}
    \end{subfigure}
    \caption{The CRT model's capability in modelling the shapes of the different components of the tagging distributions given the generator-level input of the jet flavour. The ratio bin-per-bin between the model and the target across all flavours is reported in the bottom panel.}
    \label{fig:1-d_comp_tag}
\end{figure}

\begin{figure}[H]
    \centering
    \begin{subfigure}[t]{0.45\columnwidth}
        \centering
        \includegraphics[width=\linewidth]{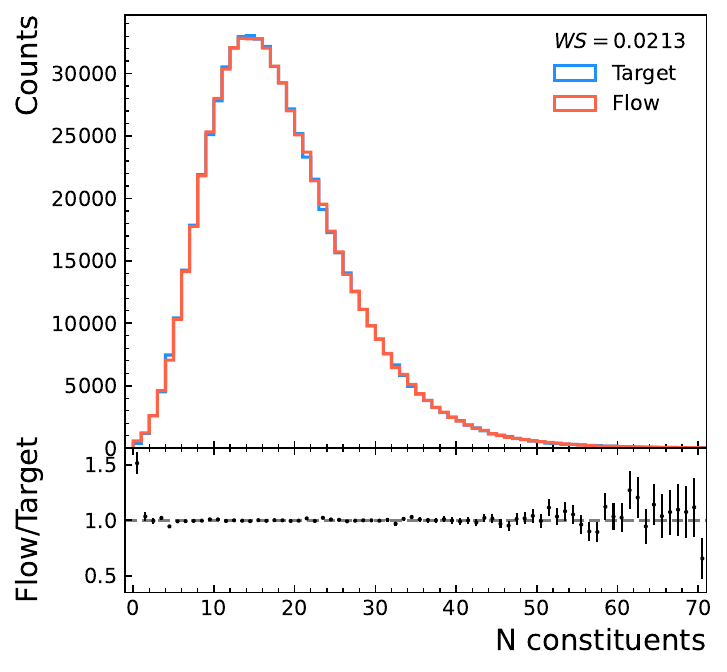}
        \caption{}
        \label{fig:ttbar_nconst}
    \end{subfigure}
    \begin{subfigure}[t]{0.45\columnwidth}
        \centering
        \includegraphics[width=\linewidth]{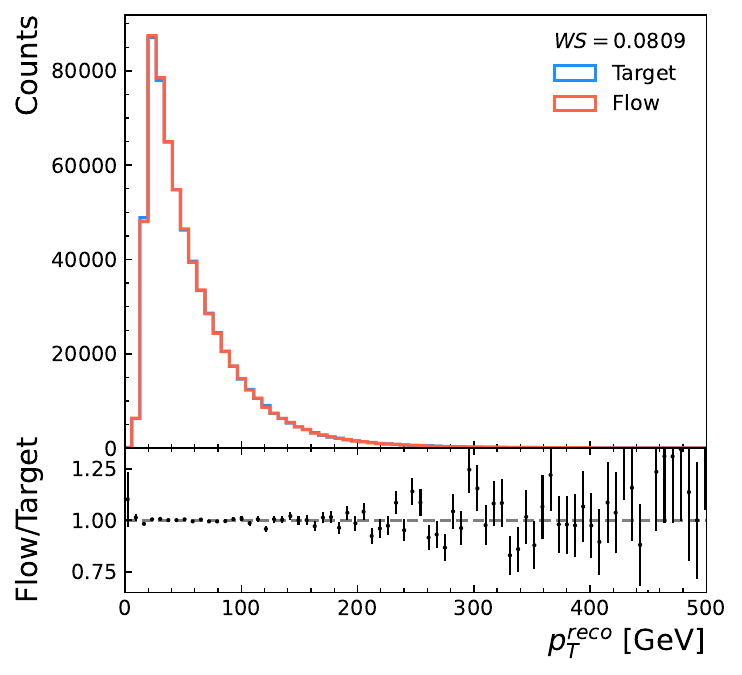}
        \caption{}
        \label{fig:ttbar_true_pt}
    \end{subfigure}
    \caption{The number of constituents (left) and the transverse momentum (right) distributions are correctly reproduced. The 1-d Wasserstein score between Target and Flow distributions is reported as a measure of convergence.}
    \label{fig:1-d_comp_phys}
\end{figure}

Figure \ref{fig:1-d_comp_tag} shows the comparison of 1-d distributions between the target dataset and the flow model results. In particular, we show the b and c-tagging distribution by input flavour: we can see that the flow is learning to correctly reproduce the different response of the tagging algorithm according to the input flavour flag. A similar level of convergence is obtained for all other distributions, such as the number of constituents or the $p_T$, showed in Figure \ref{fig:1-d_comp_phys}. The full results for 1-d distributions are shown in \ref{secA2}, Figure \ref{fig:all_distributions}.

\begin{figure}[H]
    \centering
    \includegraphics[width=\textwidth]{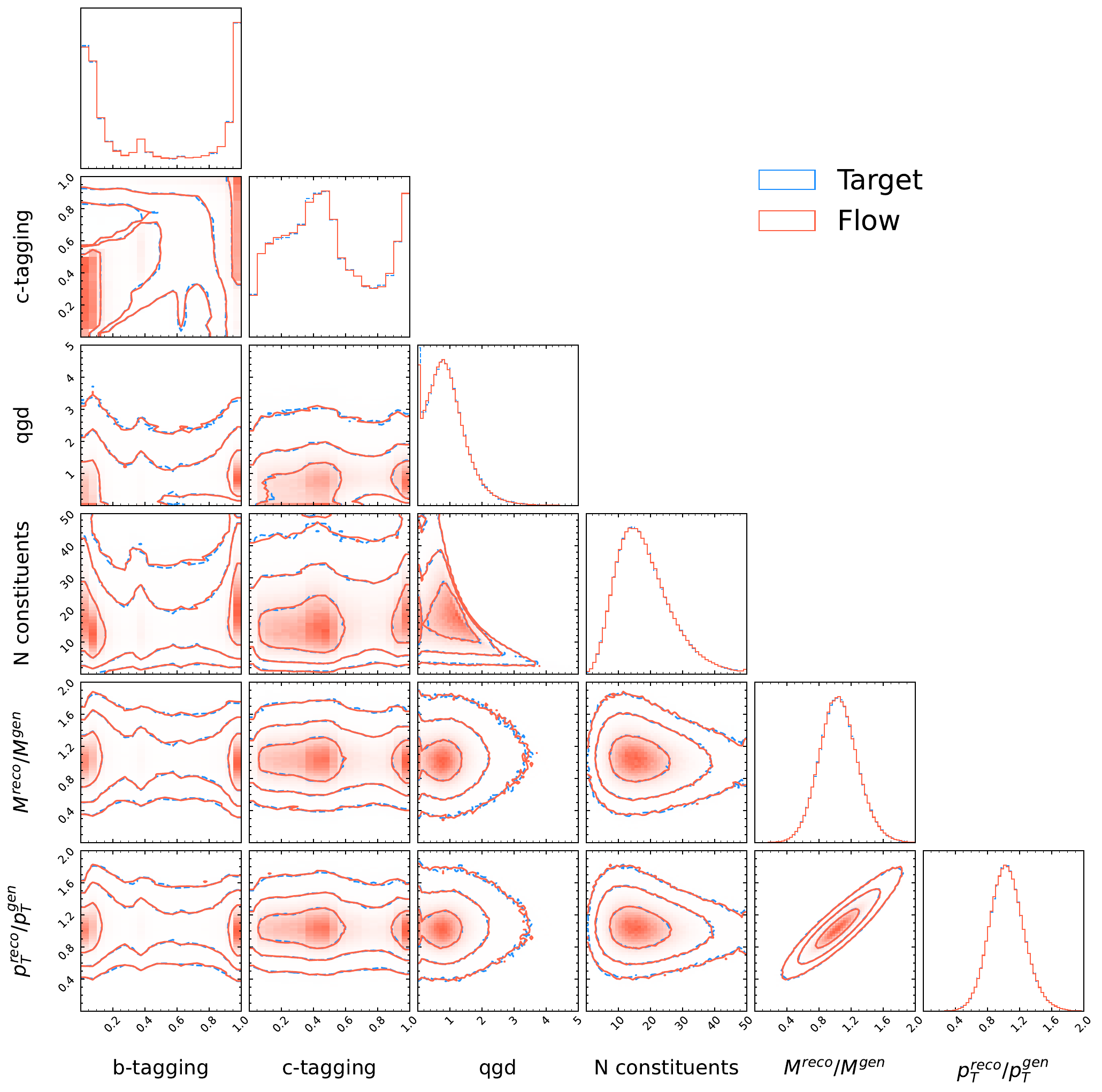}
    \caption{The correlations between different target variables are in agreement between Target and Flow. On the diagonal, we show the 1-d histograms for each variable. The off-diagonal elements compare the contour lines for 2-d distributions for each pair of variables.}
    \label{fig:corner}
\end{figure}

We show in Figure \ref{fig:corner} a plot highlighting the correlations among a subset of target variables. We can see that non-trivial correlations, such as those between the b and c-tagging algorithms, are correctly reproduced.

\clearpage

\begin{figure}[H]
    \centering
    \includegraphics[width=\linewidth]{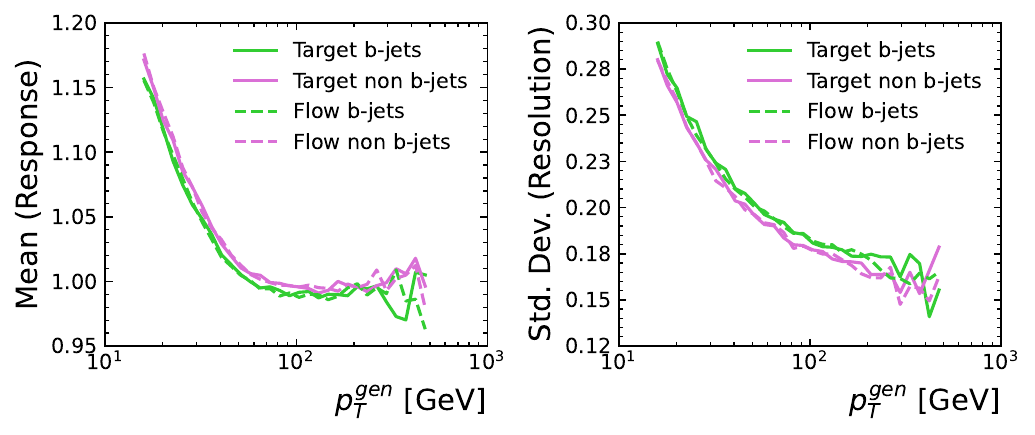}
    \caption{The plot shows the mean of the distribution (left) and its standard deviation (right) for each bin of the generator jet $p_T$. The different colours represent the different flavours, while Target vs Flow results are drawn with a different line style. We observe a general agreement on the scaling of these curves, with the subtle differences between b and non b-flavoured jets being correctly learned by the model.}
    \label{fig:reco-prof}
\end{figure}

The same is true for the correlation between the generator-level $p_T$ and the reconstructed one, shown in Figure \ref{fig:reco-prof}. The model is correctly reproducing the dependencies for both the response and resolution on jet momentum.

\begin{figure}[H]
    \centering
    \begin{subfigure}[t]{0.45\columnwidth}
        \centering
        \includegraphics[width=\linewidth]{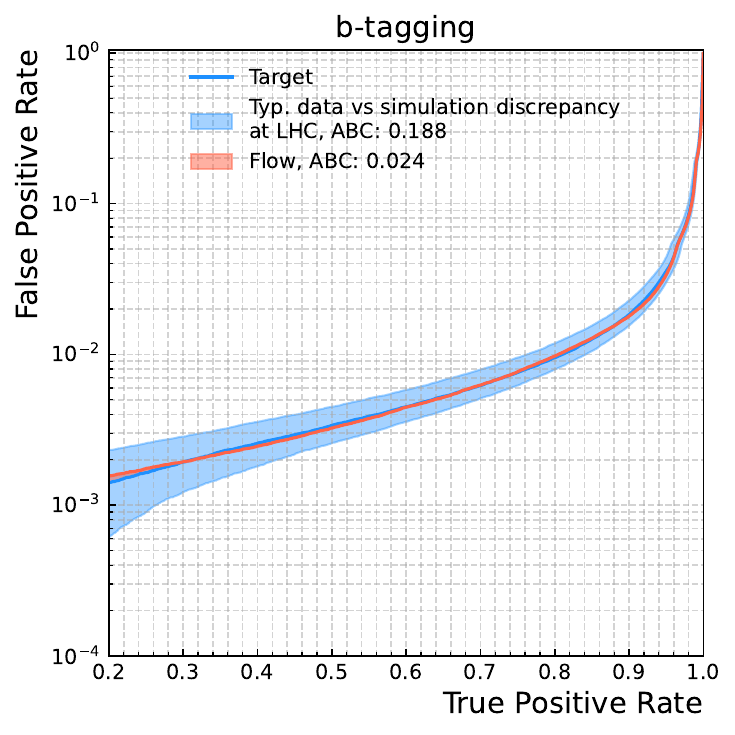}
        \caption{}
        \label{fig:ttbar_btag_roc}
    \end{subfigure}
    \hfill
    \begin{subfigure}[t]{0.45\columnwidth}
        \centering
        \includegraphics[width=\linewidth]{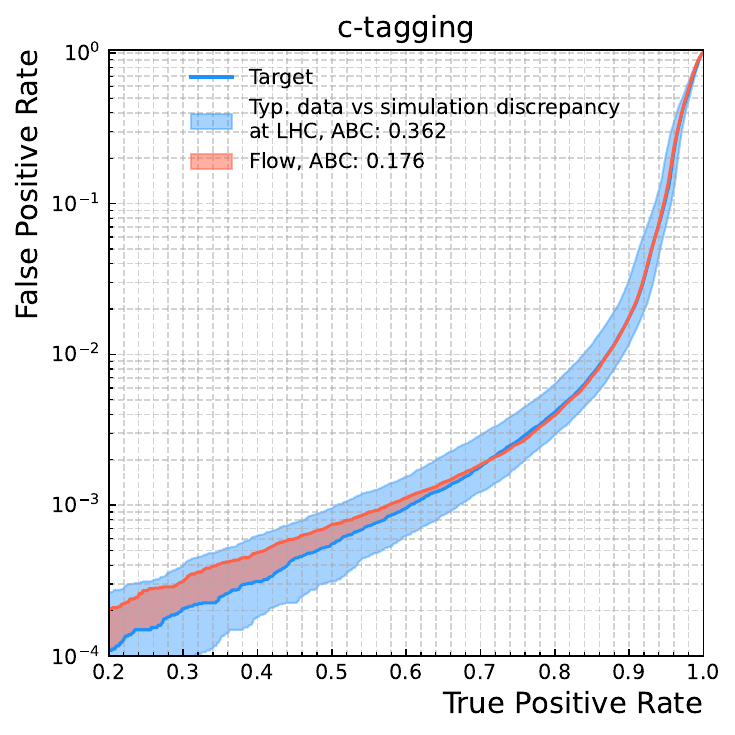}
        \caption{}
        \label{fig:ttbar_ctag_roc}
    \end{subfigure}
    \caption{ROC curves for the b (left) and c (right) tagging algorithms. The Area Between Curves (ABC) shown as the red shaded area between the two curves, is consistently smaller than the typical data vs simulation discrepancy at LHC Experiments.}
    \label{fig:rocs}
\end{figure}

A crucial figure of merit for assessing how well the model is taking into account the generator-level input is the ROC curve. It can be constructed from both the b and c-tagging distributions. While the ROC from our toy-dataset is not that of an actual experiment, it is nonetheless important to correctly reproduce it, especially in the low FPR regime (e.g. $\sim 10^{-2}$). In Figure \ref{fig:rocs} we can see the comparison between the target and model ROCs. In order to put into context, and to guide the eye, we also added a band around the target ROC, representing the typical data vs simulation differences at the LHC. The band shows $10\%$ differences in the FPR at a TPR of $\sim 50\%$ (see, e.g., the CMS experiment report \cite{Sirunyan_2018}). We can see that the Flow ROC consistently falls inside the band, with an ABC smaller than that computed between the Target ROC and one side of the band.

\subsection{Scaling of performance on training data}
\begin{figure}[H]
    \centering
    \begin{subfigure}[t]{0.45\linewidth}
        \centering
        \includegraphics[width=\linewidth]{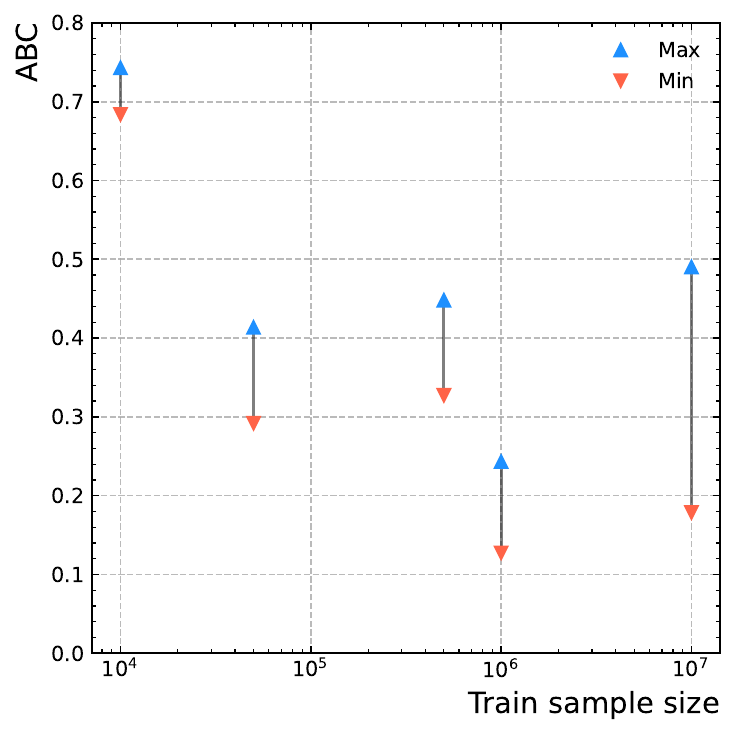}
        \caption{}
        \label{fig:qcd_ctag_roc}
    \end{subfigure}
    \begin{subfigure}[t]{0.45\linewidth}
        \centering
        \includegraphics[width=\linewidth]{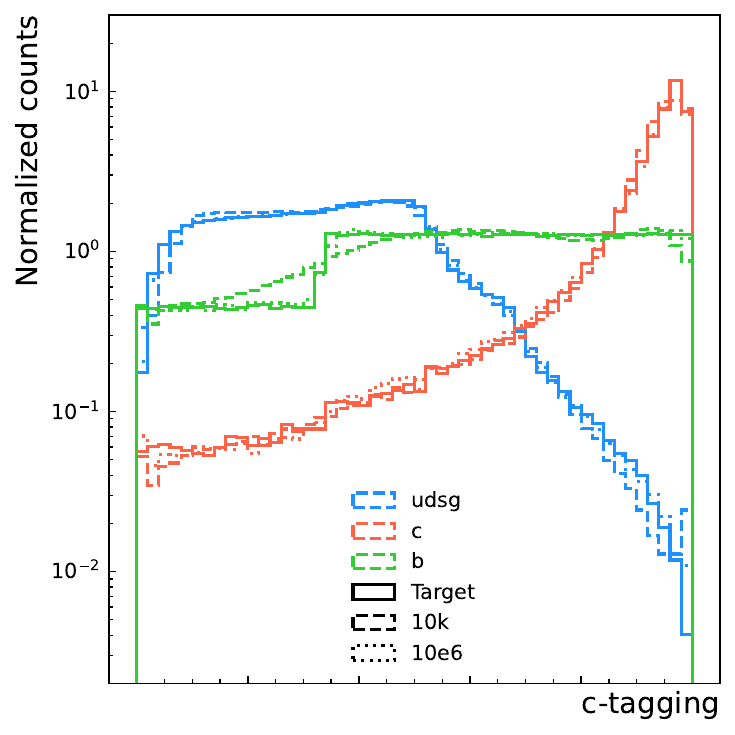}
        \caption{}
        \label{fig:zjets_ctag_roc}
    \end{subfigure}
\caption{Results when training on different data amounts. On the left, The ABC as a function of training data, with max and min values over the last 100 epochs. Despite fluctuations, we observe a downward trend. On the right, we compare the two extremes in training data. We can see that when the model is trained on a bigger dataset, the performance in the tails is modelled better than when using a smaller dataset.}
    \label{fig:training_samp}
\end{figure}

We investigated the scaling of results for the same model trained on different amounts of training data. We train the best model, CRT, on \num{10}k, \num{50}k, \num{500}k, \num{1}M and \num{10}M training samples, with a batch size of 512. We then test on a separate test split of \num{1}M events.
As expected, we observe a general improvement in results as the training split gets bigger. Figure \ref{fig:training_samp} shows how the ABC decreases as we increase the training split size. Similarly, Figure \ref{fig:training_samp} also shows how the c-tagging distribution is better modelled, especially in the tails, when using a bigger training dataset. 

We note that increasing the training dataset yields smaller and smaller gains after a certain point: the ABC performance of the model trained on 10 million samples is compatible, if not worse, with that of the model trained on 1 million.

\subsection{Performance on different physical processes}

In order to verify how well the present approach scales on different physical processes with respect to the training one, we tested the best model (trained on $t\bar{t}$) on \emph{Z+Jets}, \emph{WW} and \emph{multi-Jet} datasets never seen during training. 

\begin{figure}[H]
    \centering
        \centering
        \includegraphics[width=\linewidth]{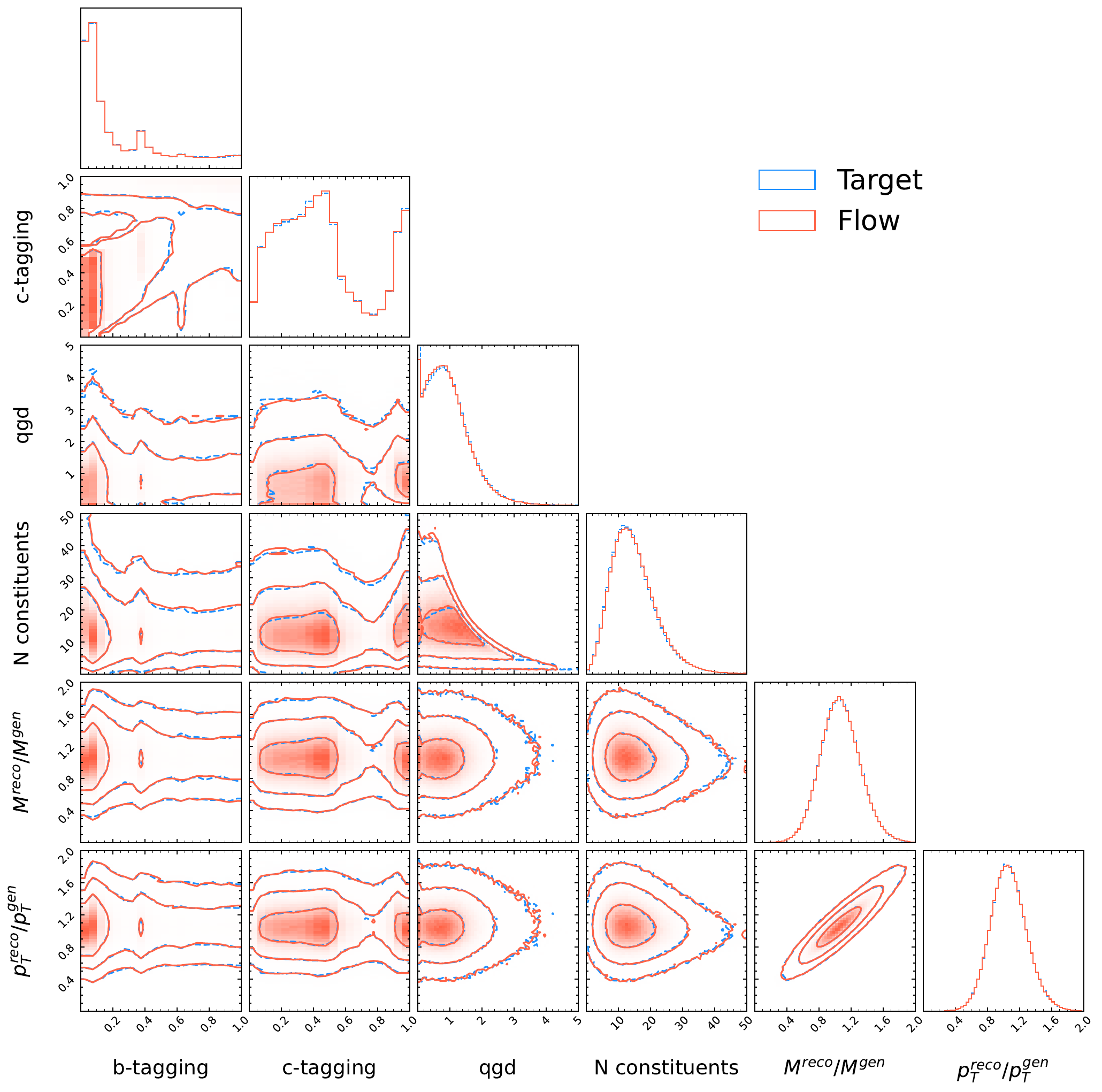}
        \label{fig:diboson_corner}
    \caption{Correlation plot for the \emph{WW} dataset, illustrating how the Target is reproduced even in physical processes different from the training ones.} 
    \label{fig:corners}
\end{figure}

\begin{figure}[H]
    \centering
    \begin{subfigure}[t]{0.45\linewidth}
        \includegraphics[width=\linewidth]{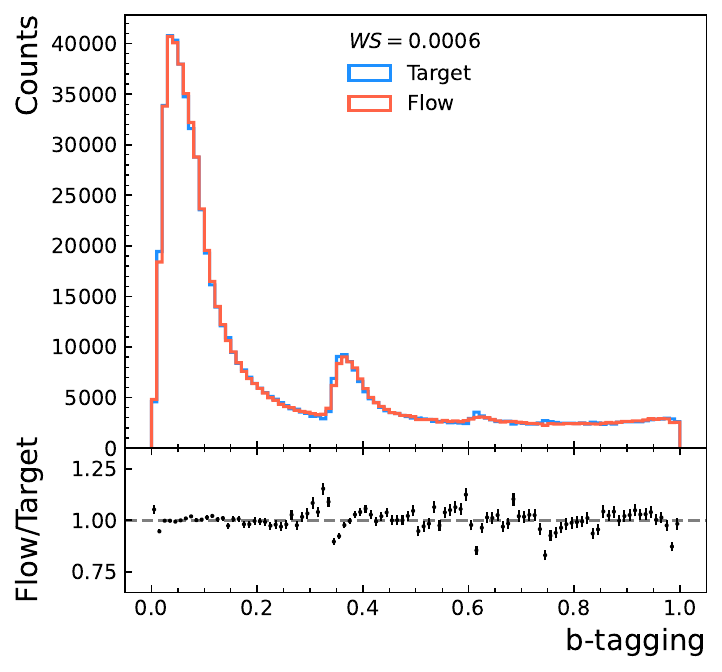}
        \caption{}
    \end{subfigure}
    \begin{subfigure}[t]{0.45\linewidth}
        \includegraphics[width=\linewidth]{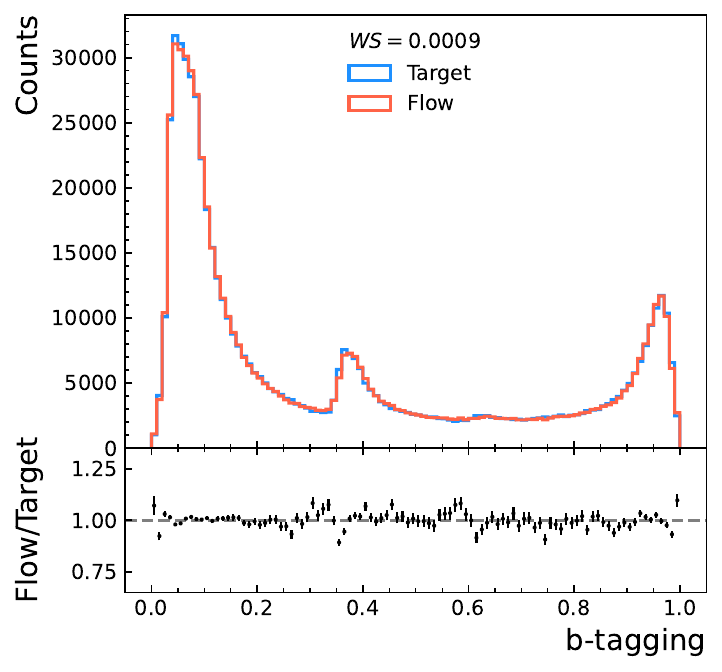}
        \caption{}
    \end{subfigure} \\
    \begin{subfigure}[t]{0.45\linewidth}
        \includegraphics[width=\linewidth]{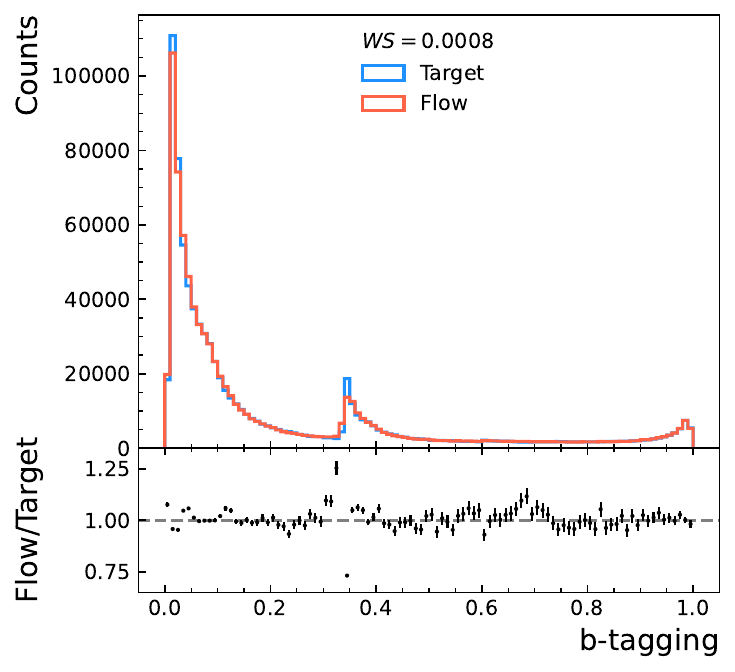}
        \caption{}
    \end{subfigure}
    \begin{subfigure}[t]{0.45\linewidth}
        \centering
        \includegraphics[width=\linewidth]{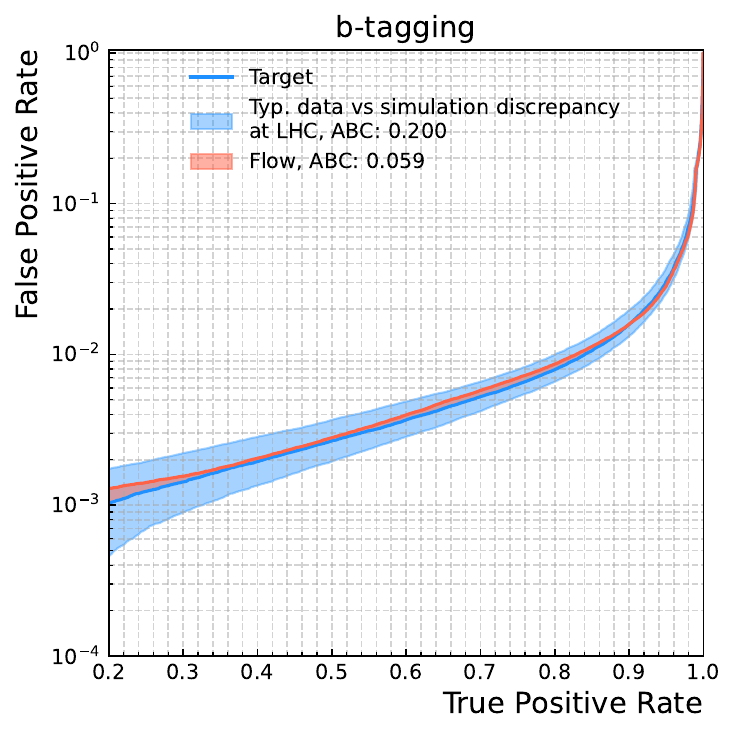}
        \caption{}
        \label{fig:qcd_btag_roc}
    \end{subfigure}
    \caption{1-d b-tagging histograms showing the performance of the model on (a) \emph{WW}, (b) \emph{Z+Jets}, (c) \emph{multi-Jet} datasets. For the latter, the associated ROC curve is shown in (d). Notice how the shapes are all different from the training process.}
    \label{fig:all-btags}
\end{figure}

The results demonstrate a good performance on different processes, as showed in Figure \ref{fig:corners}, which contains the correlation plot for the \emph{WW} dataset. Despite different tails and different correlations with respect to the physics process used in training, they are all reproduced correctly. Similar results are observed for tagging as well, as shown in Figure \ref{fig:all-btags}.

\clearpage

\clearpage

\begin{figure}[H]
    \centering
    \begin{subfigure}[t]{0.45\linewidth}
        \centering
        \includegraphics[width=\linewidth]{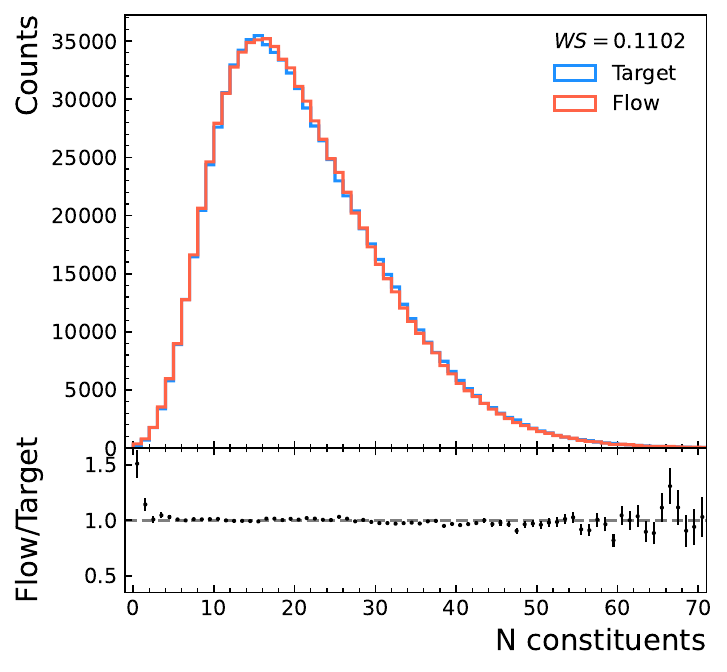}
        \caption{}
        \label{fig:diboson_nconst}
    \end{subfigure}
    \begin{subfigure}[t]{0.45\linewidth}
        \centering
        \includegraphics[width=\linewidth]{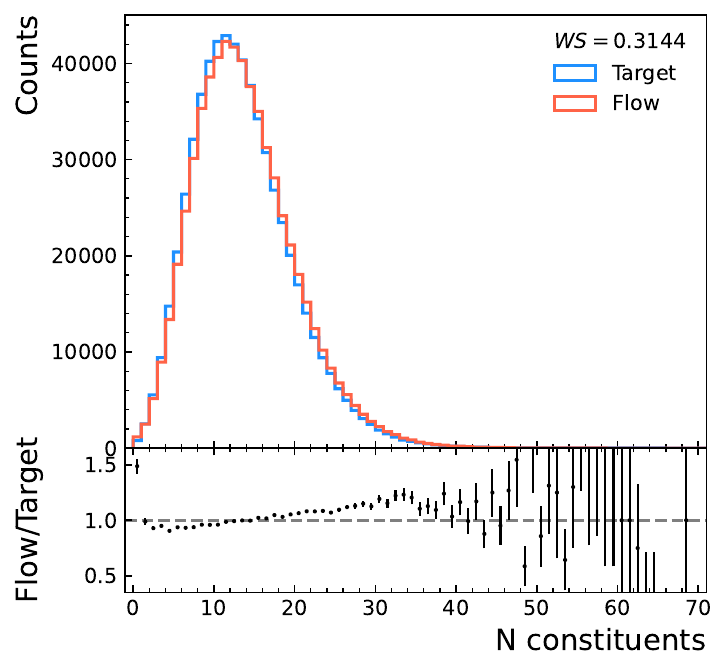}
        % \caption{\emph{Z+Jets} sample.}
        \caption{}
        \label{fig:zjets_nconst}
    \end{subfigure}
\caption{The Number of Constituents for \emph{multi-Jet} (left), and \emph{Z+Jets} (right) datasets.}
    \label{fig:more_res_proc}
\end{figure}

However, some variables show a slight bias on some datasets. As an example, Figure \ref{fig:more_res_proc} shows the number of constituents distributions, displaying a slight bias visible in the ratio plot. This is likely due to the presence of some generator-level information which is crucially correlated with this observable, but we are not giving as input to our model. For example, some jets originate from tau leptons hadronic decays and such information is missing from the model inputs, while the fraction of tau jets changes for different physical processes. A full comparison of the metrics is presented in Table \ref{tab:all_proc}, \ref{secA2}.

\subsection{Oversampling}
We demonstrated that the flow-based simulation provides a high speed and accurate detector simulation. However, because of the high simulation rate, it is interesting to study the case in which the generation step is slower compared to the end-to-end simulation (see Figure \ref{fig:problem_state}). In this case, it is convenient to use the same generator event to simulate multiple different reconstructed events. We name this procedure \emph{oversampling}. However, we must take into account the fact that reconstructed events sharing the same gen as input are correlated. Usually, in order to perform physical measurements, experimental collaborations construct probability distribution histograms for the quantities of interest in the analysis. Therefore, in our work, we propose a general procedure to build such histograms with proper bin statistical uncertainties, while considering the event correlation due to oversampling.

In particular, given a histogram, the probability associated with the $i$-th bin and its uncertainty is given by:

\begin{equation}
    p_i = \frac{\sum_{j\in\rm{bin}}w_{j}}{\sum_{k\in\rm{dataset}}w_{k}} \qquad \sigma_i = \frac{\sqrt{\sum_{j\in\rm{bin}}w^2_{j}}}{\sum_{k\in\rm{dataset}}w_{k}}
\end{equation}

where $w$ are the statistical weights associated with the events (e.g., originated by a Monte Carlo physics generator). In the oversampling case, there are $N$ events with a common gen event, called \emph{folds}, and therefore:

\begin{equation}
    p_i = \frac{\sum_{j\in\rm{bin}}\sum_{l\in\rm{fold}\in\rm{bin}}w_{jl}}{N\sum_{k\in\rm{dataset}}w_{k}} = \frac{\sum_{j\in\rm{bin}}\sum_{l\in\rm{fold}\in\rm{bin}}w_{jl} / N}{\sum_{k\in\rm{dataset}}w_{k}} \equiv \frac{\sum_{j\in\rm{bin}}w_{j}p^{\rm{fold}}_{j}}{\sum_{k\in\rm{dataset}}w_{k}}
\end{equation}

where $p^{\rm{fold}}_{j}$ is the probability associated with each fold, since each event can now enter multiple bins. Assuming that folds entering different bins are uncorrelated, the associated uncertainty becomes:

\begin{equation}
    \sigma_i =  \frac{\sqrt{\sum_{j\in\rm{bin}}(w_{j}p^{\rm{fold}}_{j})^2}}{\sum_{k\in\rm{dataset}}w_{k}}.
\end{equation}

In practice, a histogram is filled for each fold, defining $p^{\rm{fold}}_j$ as the bin content divided by $N$. The resulting histograms are then combined to obtain the final histogram. We can notice that, in this way, the resulting uncertainty for each bin is larger than the one obtained by (wrongly) considering all the events uncorrelated.

\begin{figure}[ht]
    \centering
    \includegraphics[width=0.5\linewidth]{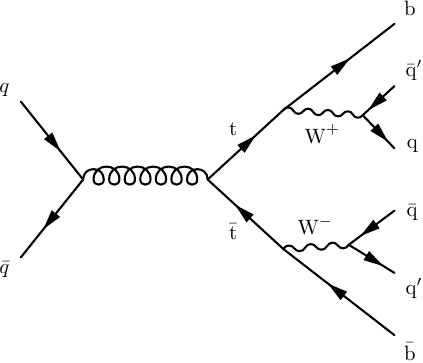}
    \caption{Feynman diagram of the $t\bar{t}$ production and subsequent decay products.}
    \label{fig:ttbar-diagram}
\end{figure}

We evaluated the effect introduced by oversampling by developing a basic analysis aiming at reconstructing a $W$ boson produced in $t\bar{t}$ process. After the production, the top quark (antiquark) decays into a $W^+(W^-)$ boson and a $b(\bar{b})$ quark. The $W$ bosons decay with higher probability into a pair of light (non-b) quarks. A general overview of the decay is shown in Figure \ref{fig:ttbar-diagram}. The resulting quarks are seen as jets of particles at the detector level.

The basic analysis works as follows: first, we selected only those jets with $p_{T} \ge 25$ GeV. Jets with b-tagging score greater than 0.5 were labelled as b-jets. Afterward, we selected the events containing precisely four jets, with two of them identified as b-tagged. Finally, we reconstructed the candidate $W$ boson from the two non-b-tagged jets. 
Oversampling has been employed to increase the number of simulated events of a \num{10}k target dataset by a factor 10. The analysis has been performed on three datasets: the original $10$k target dataset, the flow-oversampled one and a separate target dataset of $100$k events.

\begin{figure}[H]
    \centering
    \begin{subfigure}[b]{0.45\linewidth}
        \includegraphics[width=\linewidth]{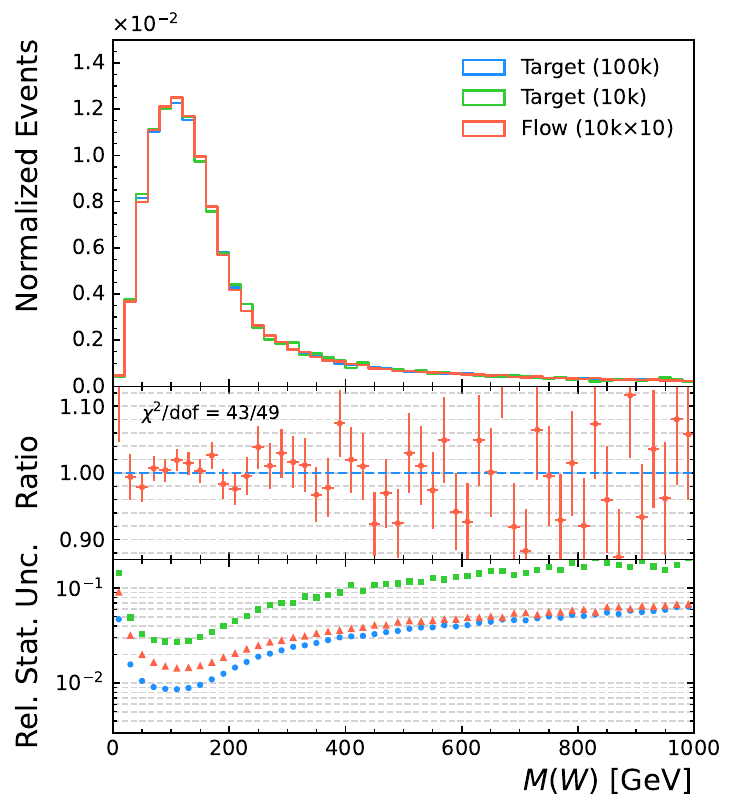}
        \caption{}
        \label{}
    \end{subfigure}
    \begin{subfigure}[b]{0.45\linewidth}
        \includegraphics[width=\linewidth]{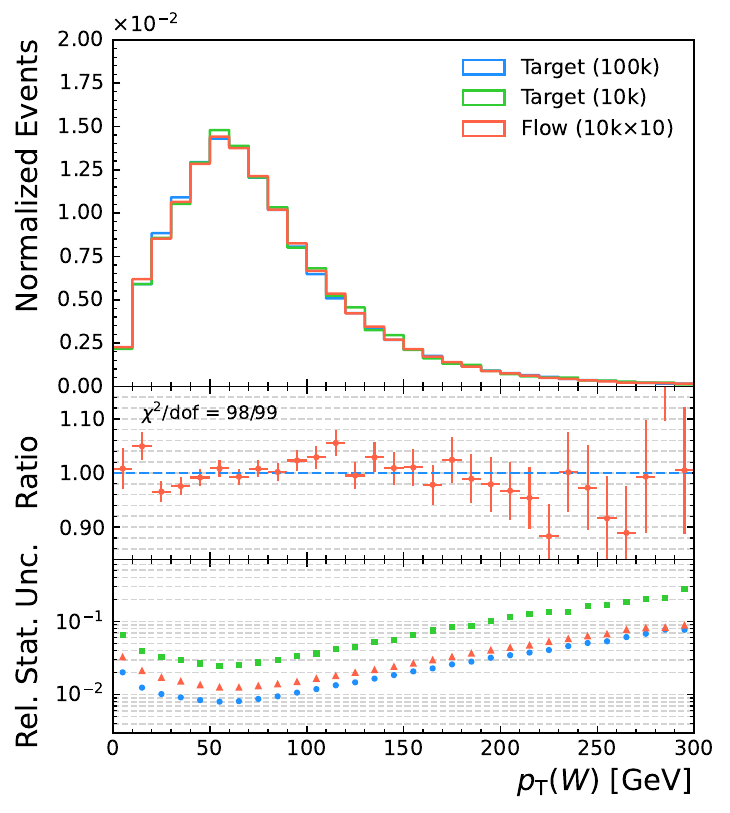}
        \caption{}
        \label{}
    \end{subfigure}\\

    \begin{subfigure}[b]{0.45\linewidth}
        \includegraphics[width=\linewidth]{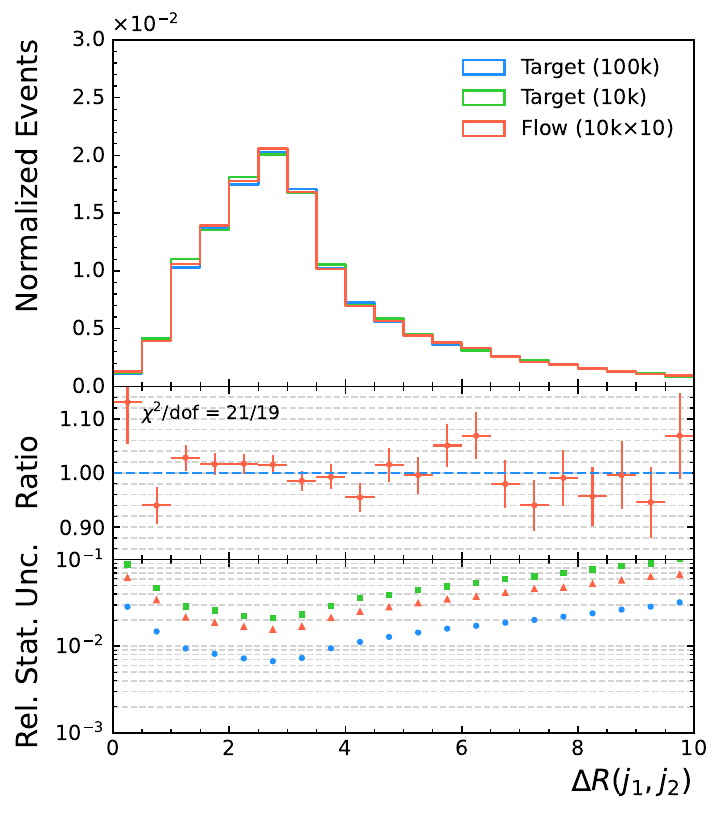}
        \caption{}
        \label{}
    \end{subfigure}
    \caption{Comparison between 100k and 10k events of the target dataset and 100k events obtained using the oversampling method for (a) invariant mass distribution of the reconstructed $W$ boson and (b) its transverse momentum. Figure (c) shows the $\Delta R$ distance between the two jets selected for the reconstruction of the $W$ boson. Each plot shows the distributions, the ratio between the oversampled dataset and its equivalent target and the relative statistical uncertainty for each bin of the histogram.}
    \label{fig:ov-w}

\end{figure}

In Figure \ref{fig:ov-w}, we show the comparison between \num{10}k and \num{100}k events of the target dataset and the \num{100}k of the oversampled dataset for the $W$ boson mass and $p_T$ distributions, as well the $\Delta R$ between the two jets used for the boson reconstruction. The $\Delta R$ measures the distance between two objects in the detector space $\eta-\phi$, and it is defined as $\Delta R = \sqrt{\Delta\eta^2 + \Delta\phi^2}$.

As expected, oversampling provides a method to increase the statistical power of the dataset by reducing  its relative statistical uncertainty. However, the improvement is larger for those distributions whose resolution is strongly dependent on detector effects. As shown in Figure \ref{fig:ov-w}, the uncertainty reduction is larger for the invariant mass and $p_T$ distributions, where the performances between the oversampled dataset and its target equivalent are compatible. The reduction is smaller for the $\Delta R$ distribution, since the angular position of the jets are measured with higher precision compared to their energy and momentum. In this case, there is little advantage in using oversampling, since there is a moderate uncertainty reduction compared to the original \num{10}k samples. Additionally, we note that there is no significant bias between the two $100$k samples.

\clearpage

\subsection{Application scenarios}\label{sec:app}

The approach presented so far can be used in different scenarios: in order to simulate datasets with new detector response from existing generated events or to create new samples, including the generation step. In the latter case, different scenarios can be imagined depending on the generation speed. In particular, the time needed for generating one event can be as low as few tens of milliseconds (as in the case of the \textsc{pythia} generator used for the toy-datasets in this work) or as high as tens of second for generators with highest theoretical accuracy. \\
The overall performance in these different tasks can be estimated as a function of the flow sampling speed (see Figure \ref{fig:timing_global}) and of the size of folds when using the oversampling technique described in the previous section.\\
We evaluated three different generation speed scenarios (0.01s/ev, 1s/ev, 20s/ev) and two cases, one with oversampling and one without. For each scenario, and with different hypothesis of flow sampling rate, we evaluated the number of events that can be produced in a day on a typical HPC node with 32 CPU core and 4 GPU (see \cite{turisini2023leonardo}) by scaling the performance observed on single GPU. In order to extrapolate from the per object sampling speed to the per event one, we considered that a typical LHC event has about 20 objects (jets, electrons, muons, etc.) to be simulated.\\
Table \ref{tab:apps} shows the results obtained. It is clear that while high sampling speed (above 10 kHz)  could be useful for the scenario where gen input already exists, a lower rate is sufficient for most of the other applications. Oversampling also plays a major role in these scenarios. 

\begin{table}
    \centering
    \caption{Comparison of millions of events produced per day on a single 4 GPU computing node in different scenarios and their ratio to a conventional simulation scenario taking 20 seconds per event. }
\lineup
\resizebox{\textwidth}{!}{
    \begin{tabular}{lcc|crrrrr|rrrrr}
    \br
    &  &     &  \multicolumn{6}{c|}{\emph{Millions of events per day on a HPC Node}} &  \multicolumn{5}{c}{\emph{Ratio to Conventional sim}}  \\
    \hline
    \bf Generator & \bf  Gen time  & \bf   Fold & \bf  Conventional & \multicolumn{5}{c|}{\bf  Object sampling speed [kHz]}  &\multicolumn{5}{c}{\bf  Object sampling speed [kHz]} \\
    & \bf  s/event  & \bf  size & \bf  (20s/event) &  \bf 1 &\bf  5 & \bf 10 & \bf 50 &\bf  100  & \bf 1 & \bf  5 &\bf  10 & \bf 50 &\bf  100  \\
    
    \hline
      Existing  & 0 & 1 & 0.138 & 17.3 & 86.4 & 172.8 & 864.0 & 1728.0  & 125 & 625 & 1250 & 6250 & 12500 \\ 
     \hline
      Simple & 0.02  & 1 & 0.138 & 15.4 & 53.2 & 76.8 & 119.2 & 128.0 & 111 & 385 & 556 & 863 & 927 \\ 
       &   & 10 & 0.138 & 17.1 & 81.3 & 153.6 & 531.7 & 768.0 & 123 & 588 & 1111 & 3847 & 5556 \\ 
     \hline
      Average  & 1  & 1 & 0.132 & 2.4 & 2.7 & 2.7 & 2.8 & 2.8 & 18 & 20 & 21 & 21 & 21 \\ 
       &   & 10 & 0.138 & 10.6 & 20.9 & 23.8 & 26.8 & 27.2 & 77 & 152 & 173 & 195 & 198 \\ 
     \hline
      Accurate & 20 & 1 & 0.069 & 0.14 & 0.14 & 0.14 & 0.14 & 0.14 & 2 & 2 & 2 & 2 & 2 \\ 
      and slow &   & 10 & 0.126 & 1.28 & 1.4 & 1.4 & 1.4 & 1.4 & 10 & 11 & 11 & 11 & 11 \\ 
     \br
    \end{tabular}}
    \label{tab:apps}

\end{table}

\section{Conclusion}\label{sec5}
We introduced Normalizing Flows for end-to-end simulation in High Energy Physics. Continuous Normalizing Flows trained with Flow Matching have been identified as the most accurate class of models, after testing on a series of physically-motivated metrics.

We demonstrated how flow-based simulation can produce accurate results for processes different from the training one, if provided with the relevant physical information. Additionally, we proposed the novel \emph{oversampling} technique, which allows using multiple reconstructed events coming from the same generator one. We showed how such a technique can effectively reduce the statistical uncertainties of existing datasets.

Moving away from the toy-datasets used in this work to an actual physics dataset, it will be important to use a series of metrics capable of measuring convergence reliably. For this task, novel goodness-of-fit testing techniques, such as the New Physics Learning Machine (NPLM) framework \cite{grosso2023goodness}, can be employed.

We also aim to investigate the use of these algorithms for the simulation of objects with no generator information (\emph{fakes}) or for global (\emph{scalar}) event quantities. Those are the missing pieces needed to perform a comprehensive end-to-end simulation of physical events.

We are excited about the potential of Flow-based approaches to aid in HEP simulation, paving the way for new discoveries.

\ack 

We would like to thank Dr. Stephen R. Green for the useful discussion and suggestions about the topic of Normalizing Flows and Flow Matching.

Research partly funded by PNRR - M4C2 - Investimento 1.3, Partenariato Esteso PE00000013 - \emph{FAIR - Future Artificial Intelligence Research} - Spoke 1 ``Human-centered AI'', funded by the European Commission under the NextGeneration EU program.\\
This work was realized using supercomputing resources and support from ICSC – Centro Nazionale di Ricerca in High Performance Computing, Big Data and Quantum Computing – and hosting entity, funded by European Union – NextGenerationEU.
This work was supported by the Open Access Publishing Fund of the Scuola Normale Superiore.


\clearpage

\appendix
\section{Toy datasets description}\label{secA1}
The dataset used for this paper has been produced using \textsc{pythia} \cite{bierlich2022comprehensive} MC generator combined with a custom Python program emulating a toy response from a hypothetical detector.
\subsection{Physics Processes}
Different physics processes are simulated with \textsc{pythia} using the same detector response. The configuration of \textsc{pythia} for the various datasets are listed in Table \ref{tab:pythia-config}.

\begin{table}[ht]
\caption{\textsc{pythia} configuration.}
\resizebox{\textwidth}{!}{
\begin{tabular}{@{}ll}
\br
\textbf{Physical process} & \textbf{Configuration} \\
\mr
 $pp\rightarrow$ & \texttt{Beams:eCM = 13000} \\
 \mr
$t\bar{t}$ & \texttt{Top:gg2ttbar = on}, \texttt{Top:qqbar2ttbar = on} \\
$Z$+Jets & \texttt{WeakBosonAndParton:qqbar2gmZg = on}, \texttt{WeakBosonAndParton:qg2gmZq = on} \\
$WW$ & \texttt{WeakDoubleBoson:ffbar2WW = on}\\
multi+Jets & \texttt{HardQCD:all = on}, \texttt{PhaseSpace:pTHatMin = 100} \\
\br
\end{tabular}
}
    \label{tab:pythia-config}
\end{table}

% Z$+Jets, $pp \rightarrow WW$ production and $pp \rightarrow JJ + X$ (QCD + Jets

Stable particles produced by \textsc{pythia} are clustered with FastJet \cite{FastJetmanualCacciari_2012}  \emph{anti-}$k_T$ jet clustering algorithm \cite{antiktCacciari_2008} with $R=0.4$.
Jets are rejected if more than 80\% of their energy is coming from a single muon or electron.
The jet flavour is established by first trying to match heavy flavour quarks (b and c) and if no heavy flavour quark is found within a radius of $\Delta R < 0.4$ jets are matched against light jets and gluons with a transverse momentum $p_T > 5$.

\subsection{Toy detector response}
\begin{itemize}
    \item Jet kinematic properties: the four-momenta of jets are represented with transverse momentum ($p_T$), pseudorapidity ($\eta$), azimuthal angle ($\phi$) and mass. The pseudo-reconstructed four-momentum is obtained by a Gaussian smearing of the corresponding generator level quantity. The smearing is larger ($\sim 10\%$) for $p_T$ and mass and smaller for the angular variable $\eta, \phi$. The response and resolution (i.e., the mean and the width of the Gaussian noise) on the $p_T$ is different depending on the jet flavour, and on the value of generated $\eta$ and $p_T$.
    \item Jet tagging variables: the scores of the tagging variables (b-tag, c-tag and quark/gluon discriminator) have an arbitrary shape obtained transforming a uniform distribution with various functions such as $\tan^{-1}(x)$ or $1-\tan^{-1}(x)$ . While typical taggers in real experiments do not have this exact shape, as they are often the output of some ML classification algorithm, the obtained shape are similar enough to those of actual experiments \cite{Sirunyan_2018}. In order to introduce some correlations between the taggers and with other variables, the scores are further biased or deformed based on a randomly generated \emph{number of secondary vertices}, Poisson distributed with a mean depending on the jet flavour, and on the jet momentum and rapidity. Finally, the b-tagger is biased if a muon is present in the jet.\\
    Quark gluon discrimination is generated transforming a uniform distribution with a power-law where the coefficient depends on the flavour of the jet. Additional correlations are introduced with the number of jet constituents and with the jet b-tagging discriminator.
    \item Number and charge of the constituents: the reconstructed number of constituents is obtained smearing the truth value and it is propagated with an additional smearing to the two variables representing the number of charged and neutral constituents. Additionally, the generator level fractions of energy in charged/neutral and hadronic/electromagnetic components are computed, without any smearing, and they are used as an additional set of targets for the Normalizing Flow. The information on number of constituents and the energy fractions are also used to simulate a \emph{jet ID} discriminating variable that has larger value for jets with more charged constituents or with higher momentum, mimicking the behaviour of the jet identification discriminators used to distinguish signal jets from jets originating from pile-up interactions or noise in actual experiments (no such jets are present in this toy simulation).
\end{itemize}

\section{Models training and comparisons}\label{secA2}

\paragraph*{Models naming convention} We use the following naming convention through the paper. The starting letter denotes the general type of flow; D: discrete, C: continuous. The second letter denotes the type of transformation for discrete flows (S: spline, A: affine) or the network type for continuous flows (R: ResNet, M: MLP). The third letter indicates the variable handling by discrete flows (C: coupling, A: autoregressive) or the flow matching algorithm for continuous flows (T: target, B: basic). For discrete flows, we also report after the name the number \emph{n} of discrete transformations applied.

\paragraph{Time prior}
Following \cite{dax2023flow}, we sample \( t \) in (7) from a power-law distribution \( p_{\alpha}(t) \propto t^{1/(1+\alpha)} \), \( t \in [0, 1] \), introducing an additional hyperparameter \( \alpha \). This includes the uniform distribution for \( \alpha = 0 \), but for \( \alpha > 0 \), assigns greater importance to the vector field for larger values of \( t \). The authors of \cite{dax2023flow} argue that a time prior \( U[0, 1] \) distributes the training capacity uniformly across \( t \) and that this is not always optimal in practice, as the complexity of the vector field may depend on \( t \).  They empirically found larger values of $\alpha$ to improve learning for distributions with sharp bounds. We found this to be true for our dataset as well, and set $\alpha =1$ in most of our continuous models training routines (and specifically in that of the best model, CRT).

\paragraph{Noise distribution} The I-CFM allows us to start from a different noise distributions rather than the typical Gaussian one. We experimented with a combination of I-CFM and a Uniform \( U[-1, 1] \) noise distribution, but we found it to have lower performances than the combination of Target flow matching and Gaussian noise source. 

\paragraph{Best model}
The detailed hyperparameters for the best model on the bigger dataset are specified in Table \ref{tab:best_hp}. 
For the ODE solver, we use torchdiffeq \cite{torchdiffeq} with solver \emph{dopri5}, \emph{atol} and \emph{rtol} both equal to $10^{-5}$ and timesteps $t=100$. 

\begin{table}[H]
\caption{Best model hyperparameters.}
\resizebox{0.6\linewidth}{!}{
\begin{tabular}{@{}ll}
\br
\textbf{Hyperparameter} & \textbf{Value} \\
\mr
Target (reco) Dimension & 16 \\
Input (gen) Dimension & 12 \\
\mr
\multicolumn{2}{c}{\textit{Training Parameters}} \\
\mr
Epochs & 1000 \\
Learning Rate & 0.001 \\
Optimizer & Adam \\
Scheduler & ReduceLROnPlateau \\
\mr
\multicolumn{2}{c}{\textit{Data Parameters}} \\
\mr
Number of Training Samples & 500000 \\
Number of Test Samples & 200000 \\
Batch Size & 64 \\
Flavour One-Hot Encoding & True \\
Standardize Data & True \\
Noise Distribution & Gaussian \\
\mr
\multicolumn{2}{c}{\textit{Model Parameters}} \\
\mr
CFM $\sigma_{\rm{min}}$ & 0.0001 \\
Matching Type & Target \\
ODE Backend & torchdiffeq, dopri5 \\
$\alpha$ & 1 \\
Timesteps & 100 \\
Type & Resnet \\
Hidden Dimensions & [32x2, 64x2, 128x2, 128x2, 64x2, 32x2] \\
Activation Function & GELU \\
Dropout & 0.0\\
Batch Normalization & False \\
Total parameters & 115440\\
\br
\end{tabular}
}
\label{tab:best_hp}
\end{table}

\paragraph*{Small dataset results} Table \ref{tab:small_comp} shows the results for models trained on the smaller dataset (6 inputs, 5 targets). We used these results to guide our choice of models to train on the bigger dataset. The two major insights drawn were:
\begin{itemize}
    \item Continuous models have a better performance than discrete ones, while using a smaller number of parameters. Their performances are quite close to each other, so we decided to retrain most of them on the bigger dataset;
    \item Among discrete models, the best ones were the DAA 20 and DAC 5 with SiLU/GELU activation functions, which were retrained on the bigger dataset.
\end{itemize}

\begin{table}[H]
\caption{The results of our experiments on the small dataset with different models. The continuous ones (starting with C) outperform all other combinations of discrete flows (starting with D).}
\resizebox{\textwidth}{!}{
\begin{tabular}{@{}lcccccc}
\br
Model & Total Parameters &  WD mean $\downarrow$& CM $\downarrow$& FGD $\downarrow$& KS Mean $\downarrow$&  ABC $\downarrow$\\
\mr
DAA, 3 & 255030  & 0.01 (12) & 0.692 (11) & 0.055 (14) & 0.14 (14) & 0.079 (4)\\
DAA, 5 & 262450  & 0.01 (12) & 1.083 (12) & 0.081 (15) & 0.128 (11) & 0.132 (7)\\
DAA, 10 & 264420  & 0.007 (9) & 2.11 (16) & 0.029 (12) & 0.146 (15) & 0.091 (5)\\
DAA, 20 & 268360  & 0.01 (12) & 1.8 (14) & 0.023 (11) & 0.125 (9) & 0.097 (6)\\
DAA, 20 & 268360 .1 & 0.096 (25) & 5.645 (29) & 3.295 (30) & 0.155 (17) & 0.396 (18)\\
DAA, 20 & 268360 .2 & 0.18 (33) & 113.976 (33) & 26.92 (33) & 0.311 (22) & 0.346 (17)\\
DAA, 20 & 268360 .3 & 0.167 (32) & 22.496 (31) & 4.045 (31) & 0.243 (21) & 1.409 (22)\\
DAA, 20 & 268360, GELU & 0.004 (7) & 0.631 (10) & 0.004 (9) & 0.12 (5) & 0.152 (8)\\
DAA, 20 & 268360, Leaky ReLU & 0.011 (14) & 0.435 (7) & 0.002 (7) & 0.127 (10) & 0.404 (19)\\
DAA, 20 & 268360, SiLU & 0.004 (7) & 0.36 (4) & 0.002 (7) & 0.136 (13) & 0.29 (16)\\
DAA, 20 & 283720, batch norm & 0.014 (16) & 3.544 (22) & 0.121 (16) & 0.156 (18) & 0.248 (14)\\
DAA, 20 & 283720, batch norm.1 & 0.013 (15) & 4.115 (27) & 0.178 (17) & 0.123 (7) & 0.253 (15)\\
DSA, 5 & 528835  & 0.101 (31) & 3.474 (18) & 0.524 (25) & 0.533 (31) & 10.934 (33)\\
DSA, 10 & 532870  & 0.099 (28) & 3.429 (17) & 0.465 (23) & 0.53 (29) & 3.915 (30)\\
DSA, 20 & 540940  & 0.092 (22) & 3.493 (19) & 0.51 (24) & 0.506 (25) & 3.901 (26)\\
DAA 10 + DSA 10 & 404650  & 0.052 (19) & 0.414 (6) & 0.006 (10) & 0.181 (19) & 1.115 (21)\\
DAC, 3 & 1315428  & 0.026 (17) & 7.339 (30) & 0.293 (18) & 0.217 (20) & 0.92 (20)\\
DAC, 5 & 1363230  & 0.009 (10) & 1.894 (15) & 0.032 (13) & 0.124 (8) & 0.203 (12)\\
DAC, 10 & 1396540  & 0.038 (18) & 1.479 (13) & 1.85 (29) & 0.149 (16) & 0.21 (13)\\
DAC, 20 & 1463160  & 0.06 (20) & 43.997 (32) & 4.77 (32) & 0.419 (23) & 1.96 (23)\\
DSC, 3 & 5433327  & 0.098 (26) & 3.885 (26) & 0.552 (26) & 0.53 (29) & 3.916 (31)\\
DSC, 5 & 5509465  & 0.092 (22) & 3.839 (25) & 0.337 (19) & 0.505 (24) & 3.897 (24)\\
DSC, 10 & 5576370  & 0.091 (21) & 4.215 (28) & 0.643 (28) & 0.509 (26) & 3.917 (32)\\
DSA, 3 & 246945  & 0.099 (28) & 3.548 (23) & 0.463 (22) & 0.545 (33) & 3.907 (27)\\
DSA, 10 & 250750  & 0.1 (30) & 3.685 (24) & 0.415 (21) & 0.525 (27) & 3.913 (29)\\
DSC, 3 & 259767  & 0.099 (28) & 3.505 (21) & 0.34 (20) & 0.534 (32) & 3.9 (25)\\
DSC, 10 & 247410  & 0.093 (24) & 3.493 (19) & 0.579 (27) & 0.53 (29) & 3.91 (28)\\
CMT & 18000& 0.003 (4) & 0.513 (9) & 0.001 (2) & 0.107 (3) & 0.17 (9)\\
CRT & 18000& 0.002 (1) & 0.448 (8) & 0.001 (2) & 0.129 (12) & 0.07 (2)\\
CRB & 18000& 0.003 (4) & 0.122 (2) & 0.001(2) & 0.112 (4) & 0.069 (1)\\
CMT small & 18000 .1& 0.003 (4) & 0.079 (1) & 0.001(2) & 0.088 (1) & 0.174 (10)\\
CMT small & 18000 .2&0.003 (4) & 0.387 (5) & 0.001 (2) & 0.102 (2) & 0.174 (10)\\
CMT bigger & 100000& 0.003 (4) & 0.122 (2) & 0.001(2) & 0.121 (6) & 0.072 (3)\\
\br
\end{tabular}
}
\label{tab:small_comp}
\end{table}

\paragraph*{Big dataset results}
We show in Table \ref{tab:gencomp} the numerical scores for the various metrics for different models. Table \ref{tab:gencomp} shows the median of the validation metrics on the last 100 epochs for each model, on the same validation split. For each metric, the ranking is shown in parentheses, where (1) stands for the best model and (10) for the worst one in the given metric. As some continuous flows are close in their results, we also compute the standard deviation for each model in each metric, and we report it under the score. In Table \ref{tab:all_proc} we show results for the best model (CRT) on different physical processes. 

We also report in the following Figure \ref{fig:all_distributions} the 1-d plots for all the target distributions of the best model on the $t\bar{t}$ test dataset.

\begin{table}[ht]
\caption{General comparison of models, where C: continuous, D: discrete. We report the median values over the last 100 epochs for a series of validation metrics. Continuous models outperform all the discrete ones. Results are quite close between continuous flows, so we report the standard deviations of each median value under it. The models are ranked across each metric, where (1) is the best and (10) the worst. The alternative simulation was obtained by computing the reported metrics on two target datasets built using the same gen input but different random seeds for the reconstruction.}
\lineup
\resizebox{\textwidth}{!}{
\begin{tabular}{lcccccc}
\br
Model &  FGD $\downarrow$& CM $\downarrow$& KS Mean $\downarrow$& WS (Scaled Mean) $\downarrow$&  ABC $\downarrow$&  c2st $\uparrow$\\
\mr
Alt. Sim. & 0.0003 & 0.1651 & 0.0955 & 0.0022 & 0.0902 & 0.5012 \\
\mr
CRT & 0.0363 (1) & 0.4214 (1) & 0.1046 (5) & 0.0042 (3) & 0.2048 (2) & 0.2573 (2)\\
 & $ 0.0009 $ & $ 0.285$ & $ 0.0957 $ & $ 0.0022 $ & $ 0.1011 $ & $ 0.002 $\\
CRT bigger & 0.0366 (2) & 0.4297 (2) & 0.1079 (7) & 0.0037 (1) & 0.1835 (1) & 0.2825 (8)\\
 & $ 0.0009 $ & $ 0.2727$ & $ 0.0958 $ & $ 0.0022 $ & $ 0.1013 $ & $ 0.0049 $\\
CRB & 0.0376 (4) & 0.6481 (6) & 0.1045 (4) & 0.0045 (5) & 0.2148 (3) & 0.2719 (5)\\
 & $ 0.0012 $ & $ 0.2944$ & $ 0.0962 $ & $ 0.0022 $ & $ 0.1009 $ & $ 0.0019 $\\
CRB alpha & 0.0369 (3) & 0.4604 (3) & 0.1075 (6) & 0.0038 (2) & 0.2691 (6) & 0.2599 (3)\\
 & $ 0.0012 $ & $ 0.2373$ & $ 0.0959 $ & $ 0.0022 $ & $ 0.1028 $ & $ 0.0017 $\\
CMB & 0.0414 (8) & 0.5861 (5) & 0.1085 (8) & 0.0064 (8) & 0.2456 (4) & 0.2768 (7)\\
 & $ 0.0013 $ & $ 0.3413$ & $ 0.0959 $ & $ 0.0022 $ & $ 0.0984 $ & $ 0.0016 $\\
CMB small & 0.0407 (7) & 1.0171 (8) & 0.1033 (3) & 0.0057 (6) & 0.36 (8) & 0.2719 (5)\\
 & $ 0.0017 $ & $ 0.4002$ & $ 0.0959 $ & $ 0.0022 $ & $ 0.0973 $ & $ 0.0021 $\\
CMB small sigma & 0.0383 (5) & 0.5636 (4) & 0.1024 (2) & 0.006 (7) & 0.299 (7) & 0.2688 (4)\\
 & $ 0.0018 $ & $ 0.3863$ & $ 0.0957 $ & $ 0.0022 $ & $ 0.108 $ & $ 0.0017 $\\
DAA gelu & 2.4281 (11) & 29.7899 (12) & 0.1148 (9) & 0.0363 (10) & 0.8154 (10) & 0.2928 (9)\\
 & $ 0.0864 $ & $ 0.6287$ & $ 0.0964 $ & $ 0.0023 $ & $ 0.1065 $ & $ 0.0064 $\\
DAA silu & 0.0803 (9) & 1.3175 (9) & 0.1645 (11) & 0.025 (9) & 0.9653 (11) & 0.312 (10)\\
 & $ 0.0063 $ & $ 0.2644$ & $ 0.0961 $ & $ 0.0022 $ & $ 0.1097 $ & $ 0.0296 $\\
DAC gelu & 3.0925 (12) & 12.0102 (11) & 0.3065 (12) & 0.1409 (12) & 1.17 (12) & 0.4514 (12)\\
 & $ 0.1411 $ & $ 0.3891$ & $ 0.096 $ & $ 0.0031 $ & $ 0.1049 $ & $ 0.0149 $\\
DAC silu & 1.5613 (10) & 8.574 (10) & 0.1187 (10) & 0.0542 (11) & 0.7417 (9) & 0.4469 (11)\\
 & $ 0.2698 $ & $ 1.0866$ & $ 0.096 $ & $ 0.0083 $ & $ 0.0963 $ & $ 0.0079 $\\
\br

\end{tabular}
}
\label{tab:gencomp}
\end{table}

\begin{table}[H]
\caption{Numerical results for the best model on the different physical processes.}
\resizebox{\textwidth}{!}{
\begin{tabular}{lcccccc}
\br
Model &  FGD $\downarrow$& CM $\downarrow$& KS Mean $\downarrow$& WS (Scaled Mean) $\downarrow$&  ABC $\downarrow$&  c2st $\uparrow$\\
\mr
zjets & 0.2028 (4) & 2.376 (3) & 0.1004 (3) & 0.012 (4) & 0.0742 (1) & 0.2455 (2)\\

diboson & 0.1466 (3) & 2.3114 (2) & 0.0951 (2) & 0.0094 (3) & 0.0757 (2) & 0.2408 (3)\\

ttbar & 0.0357 (1) & 0.3429 (1) & 0.09 (1) & 0.0037 (1) & 0.1997 (3) & 0.2552 (1)\\

qcd & 0.065 (2) & 2.424 (4) & 0.1277 (4) & 0.0059 (2) & 0.2584 (4) & 0.2371 (4)\\
 \br
\end{tabular}
}
\label{tab:all_proc}
\end{table}

\begin{figure}[H]
    \centering
    \includegraphics[width=0.24\textwidth]{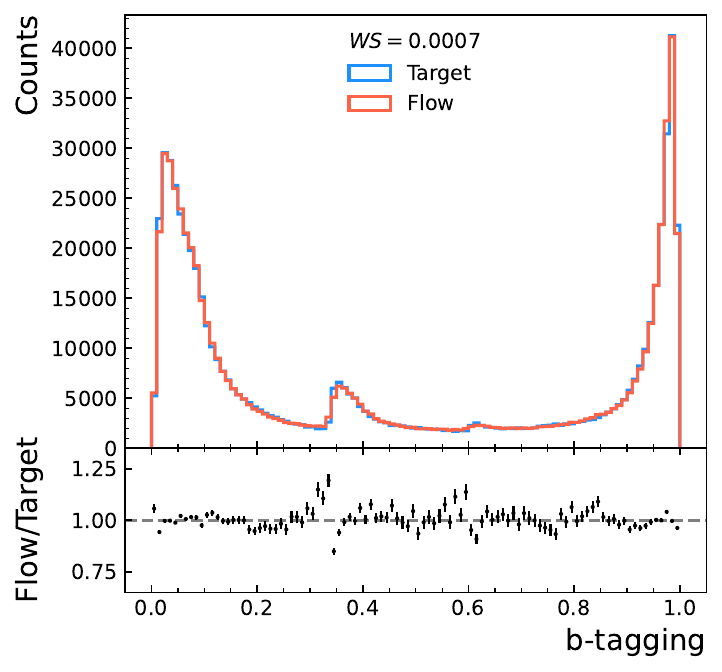}
    \includegraphics[width=0.24\textwidth]{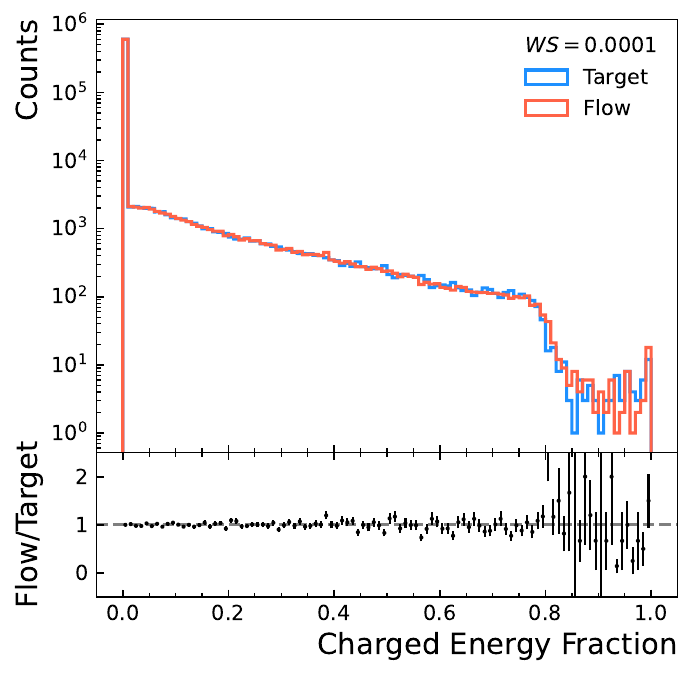}
    \includegraphics[width=0.24\textwidth]{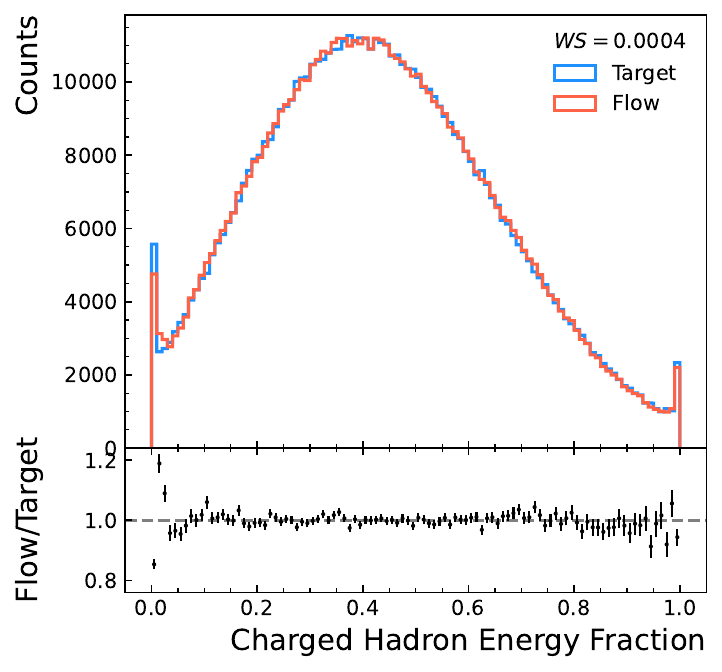}
    \includegraphics[width=0.24\textwidth]{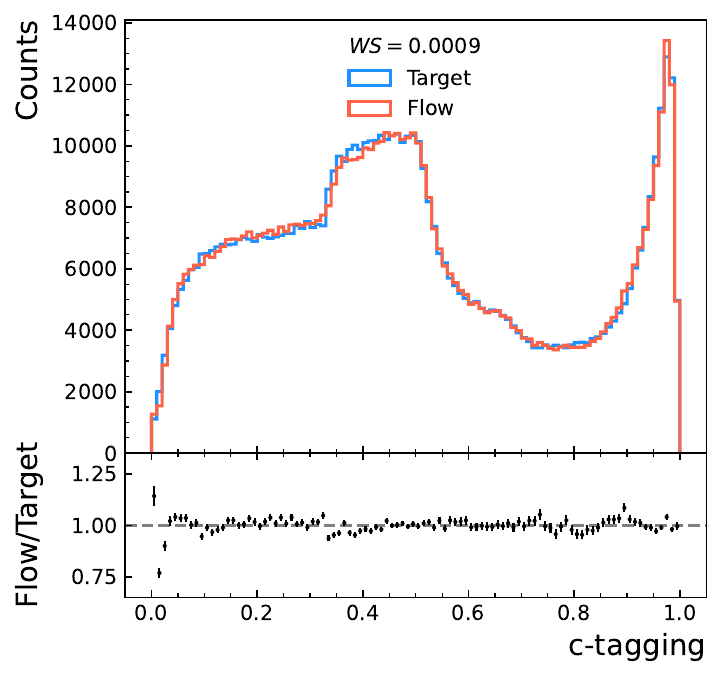}\\
    \includegraphics[width=0.24\textwidth]{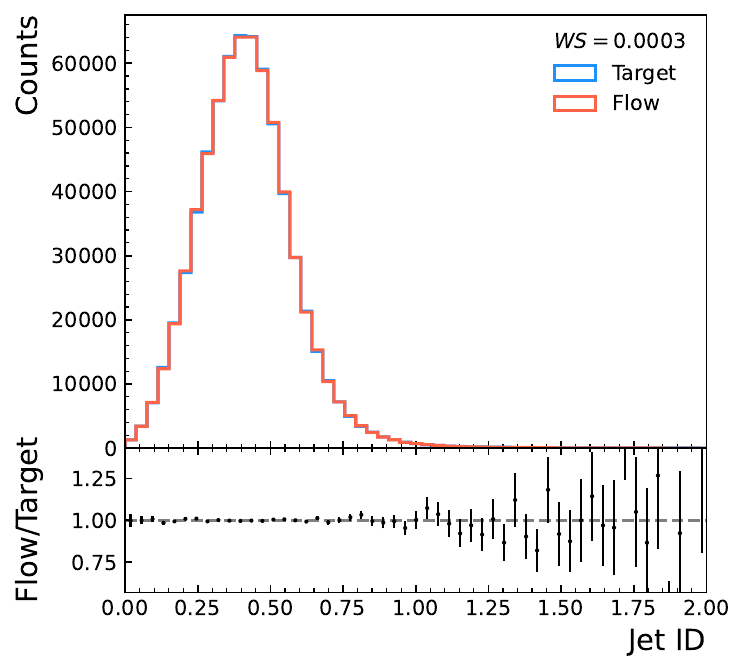}
    \includegraphics[width=0.24\textwidth]{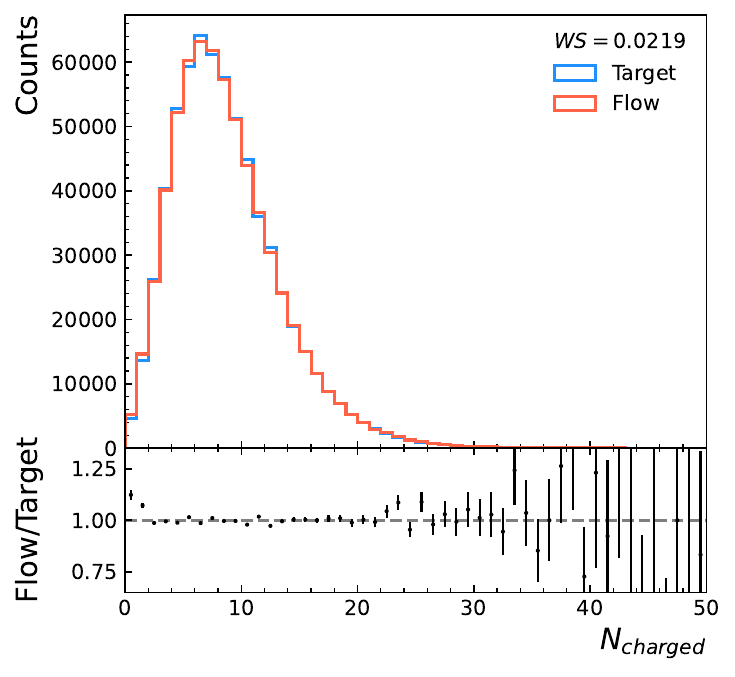}
    \includegraphics[width=0.24\textwidth]{ttbar_nconst_1d.pdf}
    \includegraphics[width=0.24\textwidth]{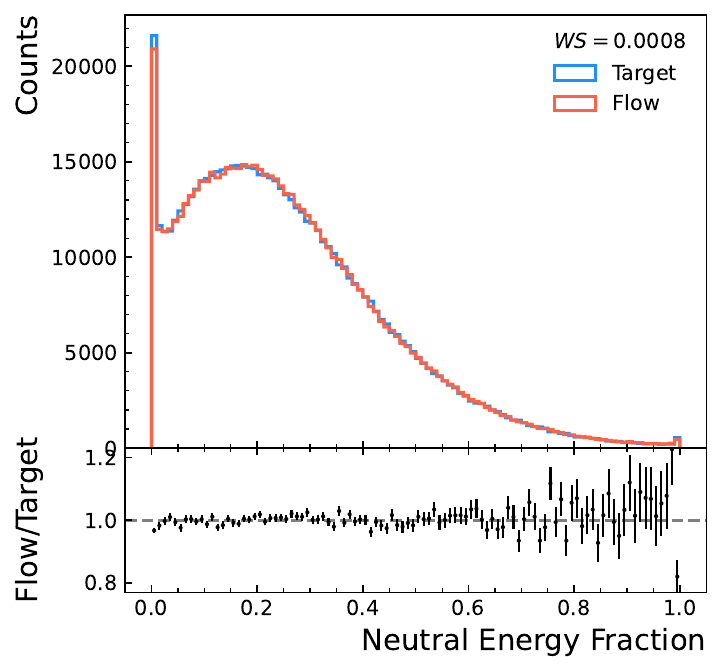}\\
    \includegraphics[width=0.24\textwidth]{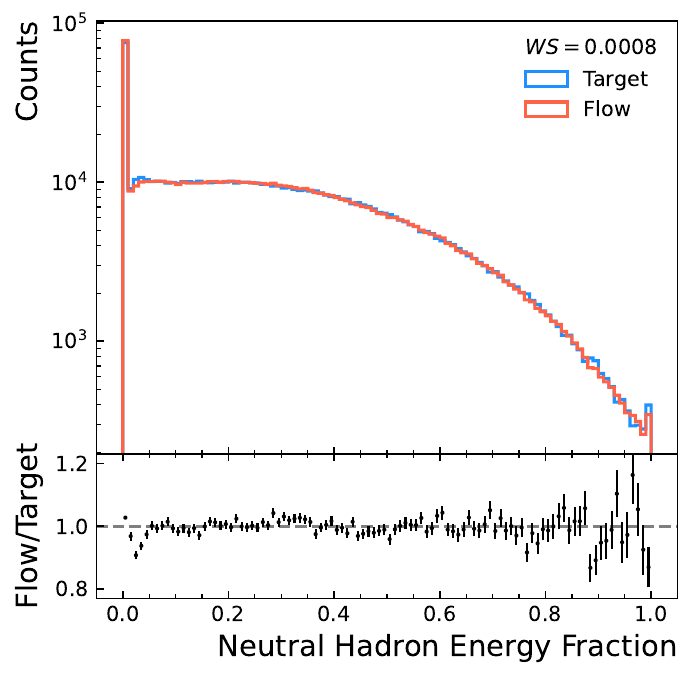}
    \includegraphics[width=0.24\textwidth]{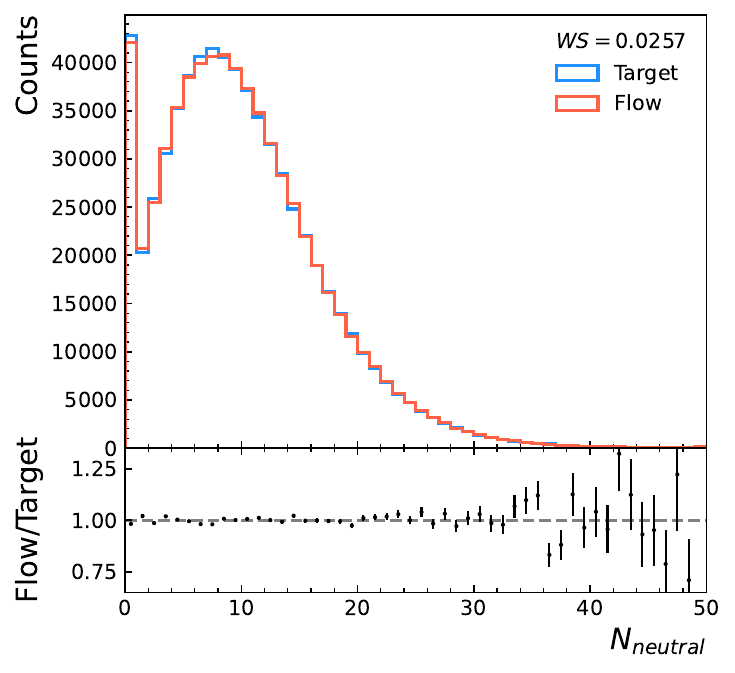}
    \includegraphics[width=0.24\textwidth]{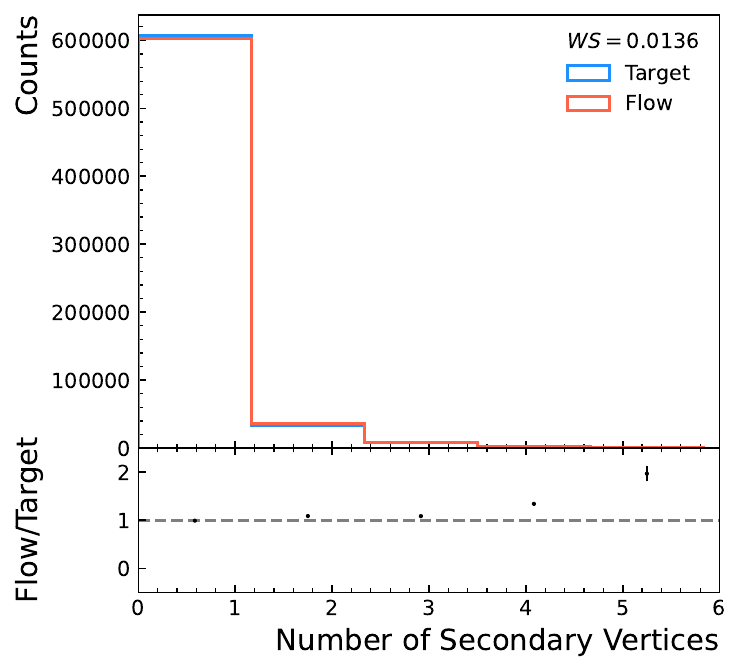}
    \includegraphics[width=0.24\textwidth]{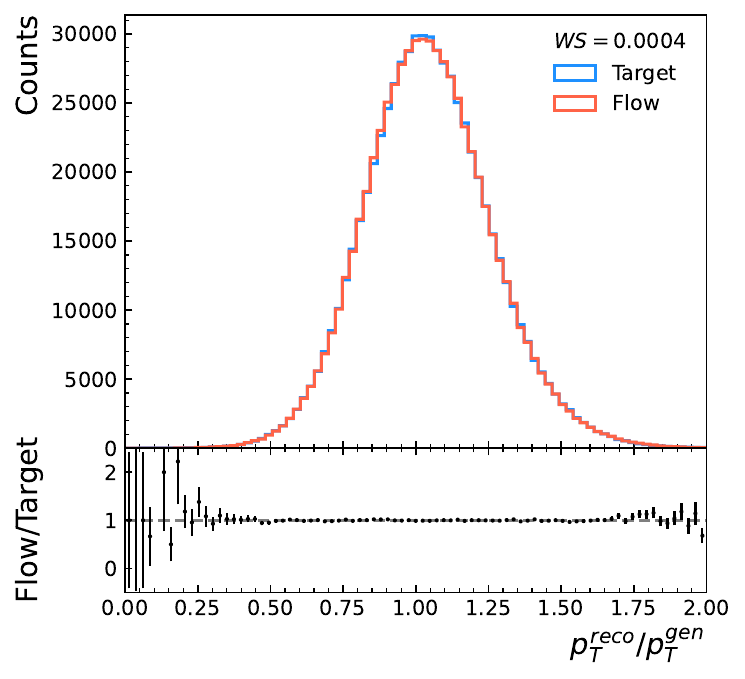}\\
    \includegraphics[width=0.24\textwidth]{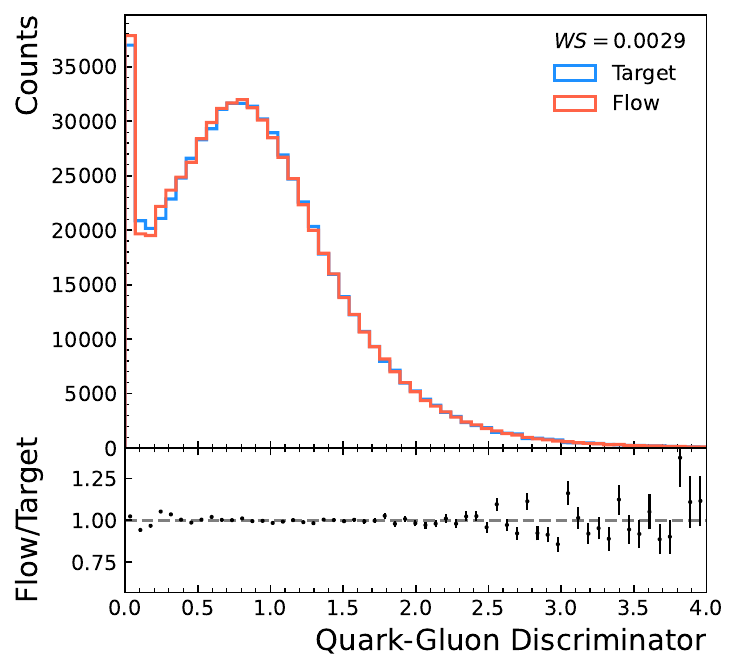}
    \includegraphics[width=0.24\textwidth]{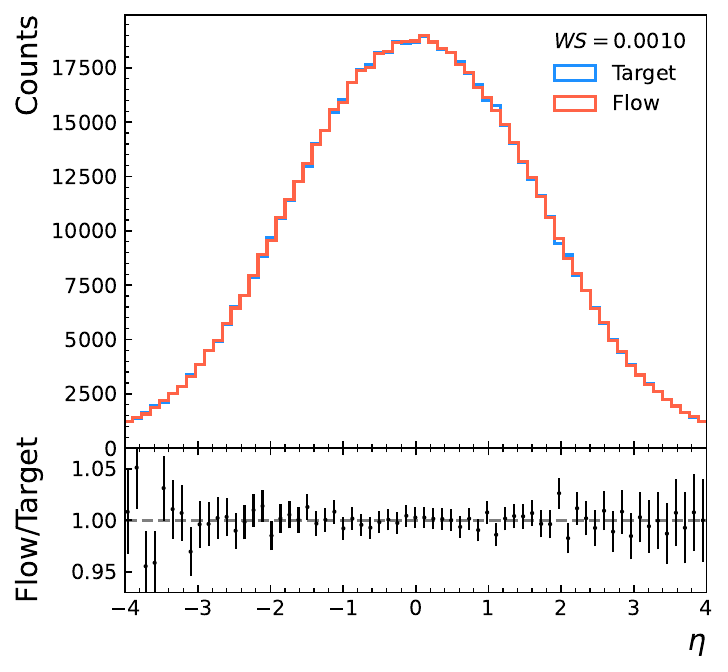}
    \includegraphics[width=0.24\textwidth]{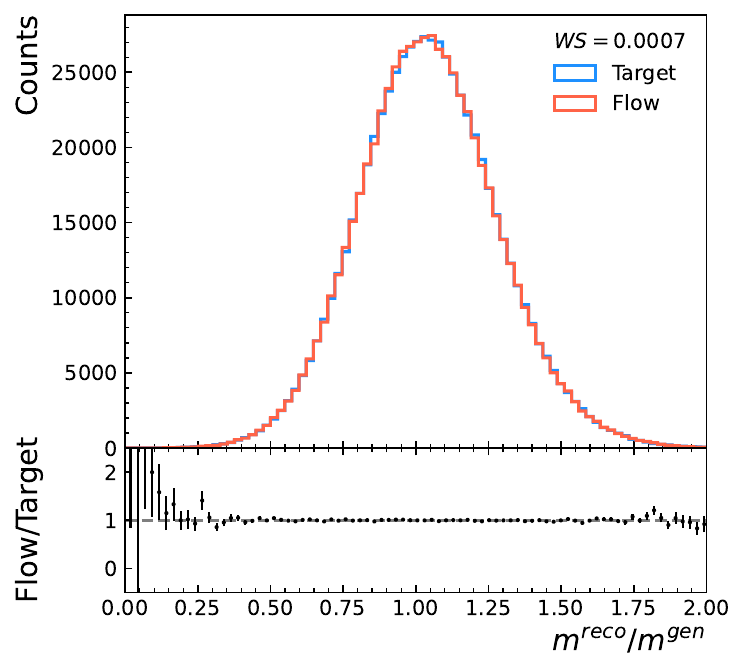}
    \includegraphics[width=0.24\textwidth]{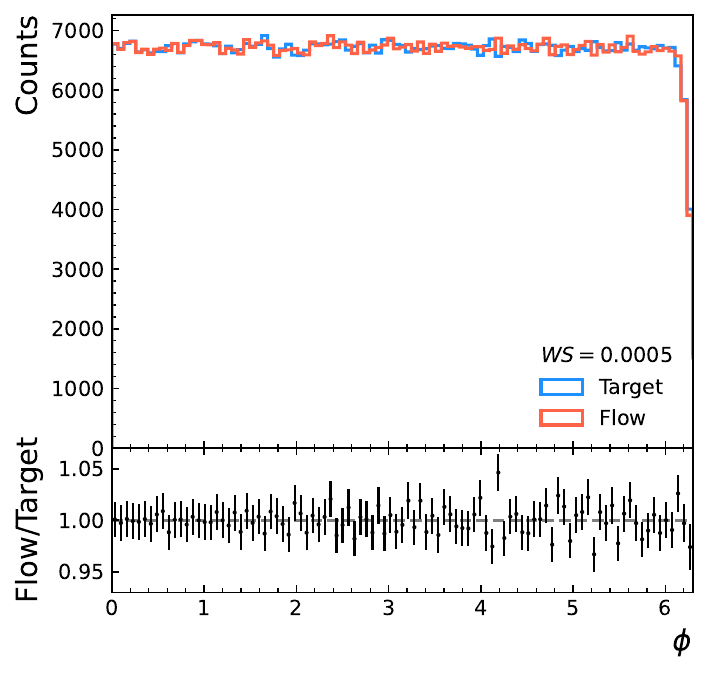}
    \caption{All the 1-d distributions obtained for the training dataset. The description of the shown variables can be found in \ref{table:datasets}.}
    \label{fig:all_distributions}
\end{figure}

\clearpage
\section*{References}
\bibliographystyle{iopart-num}
\bibliography{refs}

\providecommand{\newblock}{}
\begin{thebibliography}{10}
\expandafter\ifx\csname url\endcsname\relax
  \def\url#1{{\tt #1}}\fi
\expandafter\ifx\csname urlprefix\endcsname\relax\def\urlprefix{URL }\fi
\providecommand{\eprint}[2][]{\url{#2}}

\bibitem{Software:2815292}
Software C~O and Computing 2022 {CMS Phase-2 Computing Model: Update Document} Tech. rep. CERN Geneva \urlprefix\url{https://cds.cern.ch/record/2815292}

\bibitem{bierlich2022comprehensive}
Bierlich C, Chakraborty S, Desai N, Gellersen L, Helenius I, Ilten P, Lönnblad L, Mrenna S, Prestel S, Preuss C~T, Sjöstrand T, Skands P, Utheim M and Verheyen R 2022 {A comprehensive guide to the physics and usage of PYTHIA 8.3} (\textit{Preprint} \eprint{2203.11601})

\bibitem{AGOSTINELLI2003250}
Agostinelli S, Allison J, Amako K, Apostolakis J, Araujo H, Arce P, Asai M, Axen D, Banerjee S, Barrand G, Behner F, Bellagamba L, Boudreau J, Broglia L, Brunengo A, Burkhardt H, Chauvie S, Chuma J, Chytracek R, Cooperman G, Cosmo G, Degtyarenko P, Dell'Acqua A, Depaola G, Dietrich D, Enami R, Feliciello A, Ferguson C, Fesefeldt H, Folger G, Foppiano F, Forti A, Garelli S, Giani S, Giannitrapani R, Gibin D, {Gómez Cadenas} J, González I, {Gracia Abril} G, Greeniaus G, Greiner W, Grichine V, Grossheim A, Guatelli S, Gumplinger P, Hamatsu R, Hashimoto K, Hasui H, Heikkinen A, Howard A, Ivanchenko V, Johnson A, Jones F, Kallenbach J, Kanaya N, Kawabata M, Kawabata Y, Kawaguti M, Kelner S, Kent P, Kimura A, Kodama T, Kokoulin R, Kossov M, Kurashige H, Lamanna E, Lampén T, Lara V, Lefebure V, Lei F, Liendl M, Lockman W, Longo F, Magni S, Maire M, Medernach E, Minamimoto K, {Mora de Freitas} P, Morita Y, Murakami K, Nagamatu M, Nartallo R, Nieminen P, Nishimura T, Ohtsubo K, Okamura M, O'Neale S, Oohata Y, Paech
  K, Perl J, Pfeiffer A, Pia M, Ranjard F, Rybin A, Sadilov S, {Di Salvo} E, Santin G, Sasaki T, Savvas N, Sawada Y, Scherer S, Sei S, Sirotenko V, Smith D, Starkov N, Stoecker H, Sulkimo J, Takahata M, Tanaka S, Tcherniaev E, {Safai Tehrani} E, Tropeano M, Truscott P, Uno H, Urban L, Urban P, Verderi M, Walkden A, Wander W, Weber H, Wellisch J, Wenaus T, Williams D, Wright D, Yamada T, Yoshida H and Zschiesche D 2003 {\em Nuclear Instruments and Methods in Physics Research Section A: Accelerators, Spectrometers, Detectors and Associated Equipment\/} {\bf 506} 250--303 ISSN 0168-9002 \urlprefix\url{https://www.sciencedirect.com/science/article/pii/S0168900203013688}

\bibitem{cms_eventdisplay}
{Wikipedia contributors} 2013 {3D} view of an event recorded with the {CMS} detector in 2012 at a proton-proton centre of mass energy of 8 {TeV} [Online; accessed 8-February-2024] \urlprefix\url{https://commons.wikimedia.org/wiki/File: D_view_of_an_event_recorded_with_the_CMS_detector_in_2012_at_a_proton-proton_centre_of_mass_energy_of_8_TeV.png}

\bibitem{enwiki:1195481216}
{Wikipedia contributors} 2024 Geant4 --- {Wikipedia}{,} the free encyclopedia [Online; accessed 8-February-2024] \urlprefix\url{https://en.wikipedia.org/w/index.php?title=Geant4&oldid=1195481216}

\bibitem{de_Favereau_2014}
de~Favereau J, Delaere C, Demin P, Giammanco A, Lemaître V, Mertens A and Selvaggi M 2014 {\em Journal of High Energy Physics\/} {\bf 2014} ISSN 1029-8479 \urlprefix\url{http://dx.doi.org/10.1007/JHEP02(2014)057}

\bibitem{chen2020data}
Chen C, Cerri O, Nguyen T~Q, Vlimant J~R and Pierini M 2020 Data augmentation at the lhc through analysis-specific fast simulation with deep learning (\textit{Preprint} \eprint{2010.01835})

\bibitem{Butter_2023}
Butter A, Plehn T, Schumann S, Badger S, Caron S, Cranmer K, Di~Bello F~A, Dreyer E, Forte S, Ganguly S, Gonçalves D, Gross E, Heimel T, Heinrich G, Heinrich L, Held A, Höche S, Howard J~N, Ilten P, Isaacson J, Janßen T, Jones S, Kado M, Kagan M, Kasieczka G, Kling F, Kraml S, Krause C, Krauss F, Kröninger K, Barman R~K, Luchmann M, Magerya V, Maitre D, Malaescu B, Maltoni F, Martini T, Mattelaer O, Nachman B, Pitz S, Rojo J, Schwartz M, Shih D, Siegert F, Stegeman R, Stienen B, Thaler J, Verheyen R, Whiteson D, Winterhalder R and Zupan J 2023 {\em SciPost Physics\/} {\bf 14} ISSN 2542-4653 \urlprefix\url{http://dx.doi.org/10.21468/SciPostPhys.14.4.079}

\bibitem{Giammanco_2014}
Giammanco A 2014 {\em Journal of Physics: Conference Series\/} {\bf 513} 022012 \urlprefix\url{https://dx.doi.org/10.1088/1742-6596/513/2/022012}

\bibitem{osti_2202826}
Bein S, Connor P, Pedro K, Schleper P and Wolf M 2023  \urlprefix\url{https://www.osti.gov/biblio/2202826}

\bibitem{barbetti2023lamarr}
Barbetti M 2023 {Lamarr: LHCb ultra-fast simulation based on machine learning models deployed within Gauss} (\textit{Preprint} \eprint{2303.11428})

\bibitem{buhmann2024caloclouds}
Buhmann E, Gaede F, Kasieczka G, Korol A, Korcari W, Krüger K and McKeown P 2024 Caloclouds ii: Ultra-fast geometry-independent highly-granular calorimeter simulation (\textit{Preprint} \eprint{2309.05704})

\bibitem{ernst2023normalizing}
Ernst F, Favaro L, Krause C, Plehn T and Shih D 2023 Normalizing flows for high-dimensional detector simulations (\textit{Preprint} \eprint{2312.09290})

\bibitem{xu2023generative}
Xu A, Han S, Ju X and Wang H 2023 {Generative Machine Learning for Detector Response Modeling with a Conditional Normalizing Flow} (\textit{Preprint} \eprint{2303.10148})

\bibitem{Jawahar_2022}
Jawahar P, Aarrestad T, Chernyavskaya N, Pierini M, Wozniak K~A, Ngadiuba J, Duarte J and Tsan S 2022 {\em Frontiers in Big Data\/} {\bf 5} ISSN 2624-909X \urlprefix\url{http://dx.doi.org/10.3389/fdata.2022.803685}

\bibitem{heimel2023madnis}
Heimel T, Huetsch N, Maltoni F, Mattelaer O, Plehn T and Winterhalder R 2023 The madnis reloaded (\textit{Preprint} \eprint{2311.01548})

\bibitem{coccaro2024comparative}
Coccaro A, Letizia M, Reyes-Gonzalez H and Torre R 2024 Comparative study of coupling and autoregressive flows through robust statistical tests (\textit{Preprint} \eprint{2302.12024})

\bibitem{Mikuni_2023}
Mikuni V, Nachman B and Pettee M 2023 {\em Physical Review D\/} {\bf 108} ISSN 2470-0029 \urlprefix\url{http://dx.doi.org/10.1103/PhysRevD.108.036025}

\bibitem{Vaselli:2858890}
Vaselli F, Rizzi A, Cattafesta F and Cicconofri G (CMS) 2023 {FlashSim prototype: an end-to-end fast simulation using Normalizing Flow} Tech. rep. CERN Geneva \urlprefix\url{https://cds.cern.ch/record/2858890}

\bibitem{Krause_2023}
Krause C and Shih D 2023 {\em Physical Review D\/} {\bf 107} ISSN 2470-0029 \urlprefix\url{http://dx.doi.org/10.1103/PhysRevD.107.113003}

\bibitem{krause2023caloflow}
Krause C and Shih D 2023 Caloflow ii: Even faster and still accurate generation of calorimeter showers with normalizing flows (\textit{Preprint} \eprint{2110.11377})

\bibitem{buhmann2023epicly}
Buhmann E, Ewen C, Faroughy D~A, Golling T, Kasieczka G, Leigh M, Quétant G, Raine J~A, Sengupta D and Shih D 2023 Epic-ly fast particle cloud generation with flow-matching and diffusion (\textit{Preprint} \eprint{2310.00049})

\bibitem{Bellagente_2020}
Bellagente M, Butter A, Kasieczka G, Plehn T, Rousselot A, Winterhalder R, Ardizzone L and Köthe U 2020 {\em SciPost Physics\/} {\bf 9} ISSN 2542-4653 \urlprefix\url{http://dx.doi.org/10.21468/SciPostPhys.9.5.074}

\bibitem{birk2023flow}
Birk J, Buhmann E, Ewen C, Kasieczka G and Shih D 2023 Flow matching beyond kinematics: Generating jets with particle-id and trajectory displacement information (\textit{Preprint} \eprint{2312.00123})

\bibitem{butter2023jet}
Butter A, Huetsch N, Schweitzer S~P, Plehn T, Sorrenson P and Spinner J 2023 Jet diffusion versus jetgpt -- modern networks for the lhc (\textit{Preprint} \eprint{2305.10475})

\bibitem{Butter_2023_1}
Butter A, Heimel T, Hummerich S, Krebs T, Plehn T, Rousselot A and Vent S 2023 {\em SciPost Physics\/} {\bf 14} ISSN 2542-4653 \urlprefix\url{http://dx.doi.org/10.21468/SciPostPhys.14.4.078}

\bibitem{Gao_2020}
Gao C, Höche S, Isaacson J, Krause C and Schulz H 2020 {\em Physical Review D\/} {\bf 101} ISSN 2470-0029 \urlprefix\url{http://dx.doi.org/10.1103/PhysRevD.101.076002}

\bibitem{gavranovič2023systematic}
Gavranovič J and Kerševan B~P 2023 Systematic evaluation of generative machine learning capability to simulate distributions of observables at the large hadron collider (\textit{Preprint} \eprint{2310.08994})

\bibitem{käch2022jetflow}
Käch B, Krücker D, Melzer-Pellmann I, Scham M, Schnake S and Verney-Provatas A 2022 Jetflow: Generating jets with conditioned and mass constrained normalising flows (\textit{Preprint} \eprint{2211.13630})

\bibitem{papamakarios2021normalizing}
Papamakarios G, Nalisnick E, Rezende D~J, Mohamed S and Lakshminarayanan B 2021 {Normalizing Flows for Probabilistic Modeling and Inference} (\textit{Preprint} \eprint{1912.02762})

\bibitem{dax2023flow}
Dax M, Wildberger J, Buchholz S, Green S~R, Macke J~H and Schölkopf B 2023 {Flow Matching for Scalable Simulation-Based Inference} (\textit{Preprint} \eprint{2305.17161})

\bibitem{lipman2023flow}
Lipman Y, Chen R~T~Q, Ben-Hamu H, Nickel M and Le M 2023 Flow matching for generative modeling (\textit{Preprint} \eprint{2210.02747})

\bibitem{tong2023improving}
Tong A, Malkin N, Huguet G, Zhang Y, {Rector-Brooks} J, Fatras K, Wolf G and Bengio Y 2023 {\em arXiv preprint 2302.00482\/}

\bibitem{FastJetmanualCacciari_2012}
Cacciari M, Salam G~P and Soyez G 2012 {\em The European Physical Journal C\/} {\bf 72} ISSN 1434-6052 \urlprefix\url{http://dx.doi.org/10.1140/epjc/s10052-012-1896-2}

\bibitem{antiktCacciari_2008}
Cacciari M, Salam G~P and Soyez G 2008 {\em Journal of High Energy Physics\/} {\bf 2008} 063–063 ISSN 1029-8479 \urlprefix\url{http://dx.doi.org/10.1088/1126-6708/2008/04/063}

\bibitem{Kansal_2023}
Kansal R, Li A, Duarte J, Chernyavskaya N, Pierini M, Orzari B and Tomei T 2023 {\em Physical Review D\/} {\bf 107} ISSN 2470-0029 \urlprefix\url{http://dx.doi.org/10.1103/PhysRevD.107.076017}

\bibitem{scikit-learn}
Pedregosa F, Varoquaux G, Gramfort A, Michel V, Thirion B, Grisel O, Blondel M, Prettenhofer P, Weiss R, Dubourg V, Vanderplas J, Passos A, Cournapeau D, Brucher M, Perrot M and Duchesnay E 2011 {\em Journal of Machine Learning Research\/} {\bf 12} 2825--2830

\bibitem{paszke2019pytorch}
Paszke A, Gross S, Massa F, Lerer A, Bradbury J, Chanan G, Killeen T, Lin Z, Gimelshein N, Antiga L, Desmaison A, Köpf A, Yang E, DeVito Z, Raison M, Tejani A, Chilamkurthy S, Steiner B, Fang L, Bai J and Chintala S 2019 {PyTorch: An Imperative Style, High-Performance Deep Learning Library} (\textit{Preprint} \eprint{1912.01703})

\bibitem{tong2023simulation}
Tong A, Malkin N, Fatras K, Atanackovic L, Zhang Y, Huguet G, Wolf G and Bengio Y 2023 {\em arXiv preprint 2307.03672\/}

\bibitem{Sirunyan_2018}
CMS 2018 {\em Journal of Instrumentation\/} {\bf 13} P05011–P05011 ISSN 1748-0221 \urlprefix\url{http://dx.doi.org/10.1088/1748-0221/13/05/P05011}

\bibitem{turisini2023leonardo}
Turisini M, Amati G and Cestari M 2023 {LEONARDO}: {A Pan-European Pre-Exascale Supercomputer for HPC and AI Applications} (\textit{Preprint} \eprint{2307.16885})

\bibitem{grosso2023goodness}
Grosso G, Letizia M, Pierini M and Wulzer A 2023 Goodness of fit by {Neyman-Pearson} testing (\textit{Preprint} \eprint{2305.14137})

\bibitem{torchdiffeq}
Chen R~T~Q 2018 torchdiffeq \urlprefix\url{https://github.com/rtqichen/torchdiffeq}

\end{thebibliography}

\end{document}